\documentclass[12pt]{iopart}

\usepackage{graphicx}
\usepackage{booktabs}
\usepackage{color}
\usepackage{setstack}
\usepackage{hyperref}

\hypersetup{
pdftitle={Analysis of Lithium Driven Electron Density Peaking in FTU Liquid Lithium Limiter Experiments},
pdfauthor={Gabor Szepesi},
pdfkeywords={multi, fluid, impurity, linear, toroidal, parallel, FLR, non-trace, collisionless, electro-static, particle, transport, flux, quasi-linear},
plainpages=false,
pdfborder={0 0 0},
colorlinks=true,
linkcolor=blue,
citecolor=red,
pdfstartview={FitH},
bookmarksopen=false,
unicode
}

\newcommand{\st}[1]{\mathrm{#1}} 
\renewcommand{\vec}[1]{\mathbf{#1}}

\graphicspath{
{/home/gabor/runs/matlab_plots/ftu30582_el_transp/}
{/home/gabor/runs/matlab_plots/fluid_3sp/TEM_shat/}
{/home/gabor/runs/matlab_plots/ftu30582/nonlinear/}
{/home/gabor/data/documents/phd_work/images/ftu30582_el_transport/}
}

\begin{document}

\title[Turbulent Transport in FTU-LLL Experiments]{Analysis of Lithium Driven Electron Density Peaking in FTU Liquid Lithium Limiter Experiments}
\author{
\underline{G. Szepesi}$^{1,2}$, M. Romanelli$^2$, F. Militello$^2$, A.G. Peeters$^3$, \\ Y. Camenen$^4$, F.J. Casson$^5$, W.A. Hornsby$^3$, A.P. Snodin$^6$, \\ D. W\'agner$^7$ and the FTU team$^*$
}

\address{$^1$ Centre for Fusion, Space and Astrophysics (CFSA), University of Warwick, \\ \hspace{2mm} Coventry, CV4 7AL, UK}
\address{$^2$ Euratom/CCFE Fusion Association, Abingdon, Culham Science Centre,\\ \hspace{2mm} OX14 3DB, UK}
\address{$^3$ University of Bayreuth, Department of Physics, 95440 Bayreuth, Germany}
\address{$^4$ Aix-Marseille Universit\`e, CNRS, PIIM UMR 7345, 13397 Marseille, France}
\address{$^5$ Max-Planck-Institut f\"ur Plasmaphysik, EURATOM Association,\\ \hspace{2mm} 85748 Garching, Germany }
\address{$^6$ Department of Physics, Faculty of Science, Mahidol University,\\ \hspace{2mm} Bangkok 10400, Thailand}
\address{$^7$ Ecole Polytechnique F\'ed\'erale de Lausanne (EPFL), Centre de Recherches en \\ \hspace{2mm} Physique des Plasmas, Association Euratom-Conf\'ed\'eration Suisse, Station 13, \\ \hspace{2mm} CH-1015 Lausanne, Switzerland}
\address{$^*$ See Appendix of A.A. Tuccillo et al., OV/4-2, Fusion Energy 2010 \\ \hspace{2mm} (Proc. 23rd  Int. Conf. Daejon) IAEA (2010).}

\ead{g.szepesi@warwick.ac.uk}

\begin{abstract}
The impact of lithium impurities on the microstability and turbulent transport characteristics in the core of a typical FTU Liquid Lithium Limiter (LLL) (Mazzitelli et al., Nucl. Fusion, 2011) discharge during the density ramp-up phase is studied. 
A non-linear gyrokinetic analysis performed with GKW (Peeters et al., Comp. Phys. Comm., 2009) accompanied by a quasi-linear fluid analysis is presented. 
We show that a centrally peaked, high concentration lithium profile contributes to the electron peaking by reducing the outward electron flux, and that it leads to inward turbulent deuterium transport through ion flux separation.
\end{abstract}

\maketitle

\section{Introduction}

Understanding the driving mechanisms of particle flux in tokamaks is of major importance in addressing the problem of fuelling a tokamak reactor and determining the particle confinement capabilities of future fusion devices. Fusion power is proportional to the density of the fuel ions (that is, linear in both deuterium and tritium densities), a peaked density profile of these ions is therefore highly desired. At the same time, accumulation of impurities in the plasma core would lead to a deterioration of energy confinement due to radiative losses and significant dilution. It has been the focus of both experimental and theoretical efforts to find operating regimes which provide inward main ion and outward impurity particle flux. 

It has been pointed out by a number of authors that the effective turbulent particle transport in magnetized fusion plasmas is determined by a delicate balance between several contributing factors \cite{angioni_rev,bourdelle_rev,bourdelle_qualikiz}. The particle flux can be formally separated into a diffusive part explicitly proportional to the density gradient, a thermo-diffusive part proportional to the temperature gradient, and a residual term often called the particle pinch \cite{angioni_rev}. 
In rotating plasmas an additional term, the rotodiffusion is also taken into account \cite{camenen2009, angioni_nf_2011}. 
These contributions have been analysed in detail by both quasi-linear fluid and gyrokinetic methods in terms of their dependences on magnetic shear \cite{nordman, nordman_imp, moradi, kinsey}, magnetic curvature \cite{angioni_rev, frojdh}, collisionality \cite{angioni_coll,angioni_coll_prl}, impurity concentration and charge number \cite{nordman_imp, moradi, frojdh, angioni_prl, emila, skyman, dubuit, fulop}, and whether the dominant instability driving the small scale turbulence is an ion or electron mode \cite{angioni_rev,fable,frojdh_jarmen,angioni_ECH}. 

In this paper we use the theoretical framework introduced in the publications above in order to understand the recent experimental observations on the Frascati Tokamak Upgrade (FTU) following the installation of a Liquid Lithium Limiter (LLL): discharges performed with LLL exhibit significantly increased electron confinement properties and density peaking compared to the standard metallic limiter scenarios \cite{mazzitelli}. Similar observations have been made in various other tokamaks \cite{murakami,mckee,weynants}.

One of the possible explanations for this behaviour is that the initial high concentration of lithium impurities alters the turbulent mode spectrum and the phase difference between fluctuations and thus leads to an inward anomalous particle transport, in contrast with the metallic limiter operation (without LLL). 
A linear gyrokinetic analysis of the FTU-LLL discharge \#30582 performed with GKW \cite{gkw} using a Lorentz-type collision operator (pitch-angle scattering only) was published by Romanelli et al. \cite{ftu_lin}.
They obtained the stabilizing effect of the lithium ions on the main ITG modes due to dilution \cite{frojdh} and on ETG modes due to more effective screening of the potential fluctuations \cite{reshko}, and destabilization of the impurity ITG modes. 
They found that in the linear phase the deuterium ITG modes drive inward deuterium, electron and outward impurity flux at the early stage of the discharge. However, the direction of the transport remained unchanged when varying the density, temperature and temperature gradient of the species, and, most importantly, impurity concentration. 
It was therefore difficult to clarify the role of the lithium concentration in reversing the deuterium flux.

In this paper we extend the linear analysis of \cite{ftu_lin} to include scans of collisionality and impurity density gradient, and compare non-linear simulations with experimentally measured fluxes. 
Deuterium flux reversal between low and high impurity concentration cases in the linear simulations is found when the full collision operator is taken into account in the simulations.
We also show that a centrally peaked, high concentration lithium profile is in fact responsible for the inward turbulent deuterium flux at the startup of the discharge, and that it also contributes to the electron density peaking by reducing the outward electron flux.  
A study by Estrada-Mila et al. \cite{emila} has been investigating particle transport in a standard case using helium and tritium impurities with non-linear gyrokinetic analysis. We apply a similar approach for the analysis of this FTU discharge.
The gyrokinetic analysis is complemented by a quasi-linear fluid model that contains the necessary physics needed to capture the main aspects of the observed particle transport. The fluid approach allows us to analyse all the eigenmodes of the system and estimate their diffusive, thermo-diffusive and pinch contributions to the particle flux separately. The validity of the linear and quasi-linear analysis is confirmed by a set of non-linear gyrokinetic simulations.

The structure of the paper is as follows: In section \ref{sec:gk_lin} the extended linear gyrokinetic analysis is presented and the key parameters leading to deuterium flux reversal are identified. Non-linear simulation results supporting the main linear conclusions are also shown. In section \ref{sec:fluid} the fluid model is introduced and the results compared with those of the gyrokinetic analysis. Conclusions are summarized in section \ref{sec:conclusions}.

\section{Gyrokinetic Analysis} \label{sec:gk_lin}  

\subsection{Linear Analysis}

The simulations have been carried out with GKW \cite{gkw}, a local, initial value, Eulerian gyrokinetic code. It uses a linearized Fokker--Planck collision operator consisting of pitch-angle scattering, energy scattering and friction terms \cite{gkw}. The exact form of the collision operator is outlined in \cite{karney}. Energy and momentum conserving terms have not been included for these simulations since they are only important for neoclassical studies. 
Since FTU is not fitted with a Neutral Beam Injection (NBI) system, the plasma rotation is typically of the order of the diamagnetic velocity. 
At this rotation speed centrifugal effects (implemented in GKW \cite{Casson2010}) are not expected to affect the main ion and light impurity transport, and are not included in the present analysis.

The simulations have been performed in the good confinement region at the radius of $r/a=0.6$. Two time instances are investigated, one at the startup of the discharge during the transient ramp up phase ($t=0.3\st{s}$) characterized by $Z_{\st{eff}} = n_{\st{e}}^{-1} \sum_{\st{s}} n_{\st{s}} Z_{\st{s}}^2 = 1.93$ due to the high lithium concentration ($c_{\st{Li}}=n_{\st{Li}}/n_{\st{e}}=0.15$), and one in the density plateau phase at $t=0.8\st{s}$ with $c_{\st{Li}}=0.01$ and $Z_{\st{eff}} = 1.06$. Figure 1. in \cite{ftu_lin} shows the time evolution of the electron current, core and edge electron density and gas fuelling rate, as well as the electron density and temperature profiles at the two time instances. 

The input parameters of the simulations are the same as in \cite{ftu_lin}, reported here in table \ref{tab:parameters}. 
The reference density is chosen to be the electron density, and the reference temperature the ion temperature. The thermal velocity is defined as $v_{\st{th,s}} = \sqrt{2 T_{\st{s}} / m_{\st{s}}}$. The reference Larmor radius is the ion thermal Larmor radius on the magnetic axis: $\rho_{\st{ref}}=\rho_{\st{i}} = (m_{\st{i}} v_{\st{th,i}}) / (e B_{\st{ref}})$. The linear simulations have been performed using 264 points along the parallel direction covering 11 periods around the torus, 10 points in magnetic moment and 36 points in parallel velocity space. A non-shifted circular geometry described in \cite{lapillone} has been applied. 

\begin{table}
\begin{center}
\begin{tabular}{c c c c c}
 & $n [10^{19} \mathrm{m}^{-3}]$ & $T [\mathrm{keV}]$ & $-\frac{a \nabla T}{T}$ & $-\frac{a \nabla n}{n}$ \\
 \cmidrule{2-5}
 & \multicolumn{4}{c}{\textcolor{red}{$t=0.3\mathrm{s}$}} \\
 \midrule
 $D^+$ & 3.21 & 0.47 & 2.14 & 0.89 \\
 \midrule
 $e^-$ & 6.03 & 0.57 & 3.07 & 0.89 \\
 \midrule
 $Li^{3+}$ & 0.94 & 0.47 & 2.14 & 0.89 \\
 \midrule
 & $R_{\st{ref}}=0.97 \st{m}$ & $q=2.76$ & $\hat{s}=0.97$ & $\nu_{\st{ii,N}}=0.028$ \\
 \cmidrule{2-5}
 & \multicolumn{4}{c}{\textcolor{magenta}{$t=0.8\mathrm{s}$}} \\
 \midrule
 $D^+$ & 15.50 & 0.26 & 4.61 & 3.12 \\
 \midrule
 $e^-$ & 15.95 & 0.24 & 4.94 & 3.12 \\
 \midrule
 $Li^{3+}$ & 0.15 & 0.26 & 4.61 & 3.12 \\
 \midrule
 & $R_{\st{ref}}=0.98 \st{m}$ & $q=2.36$ & $\hat{s}=1.55$ & $\nu_{\st{ii,N}}=0.220$ \\
 \bottomrule
\end{tabular}
\caption{Plasma parameters used in the gyrokinetic simulations at $r/a = 0.6$, $t_1=0.3\ \mathrm{s}$ and $t_2=0.8\ \mathrm{s}$. $a=0.3\ \mathrm{m}$.}
\label{tab:parameters}
\end{center}
\end{table}

\subsubsection{The Density Ramp-up Phase}

\begin{figure}
 \begin{center}
  \includegraphics[scale=0.25]{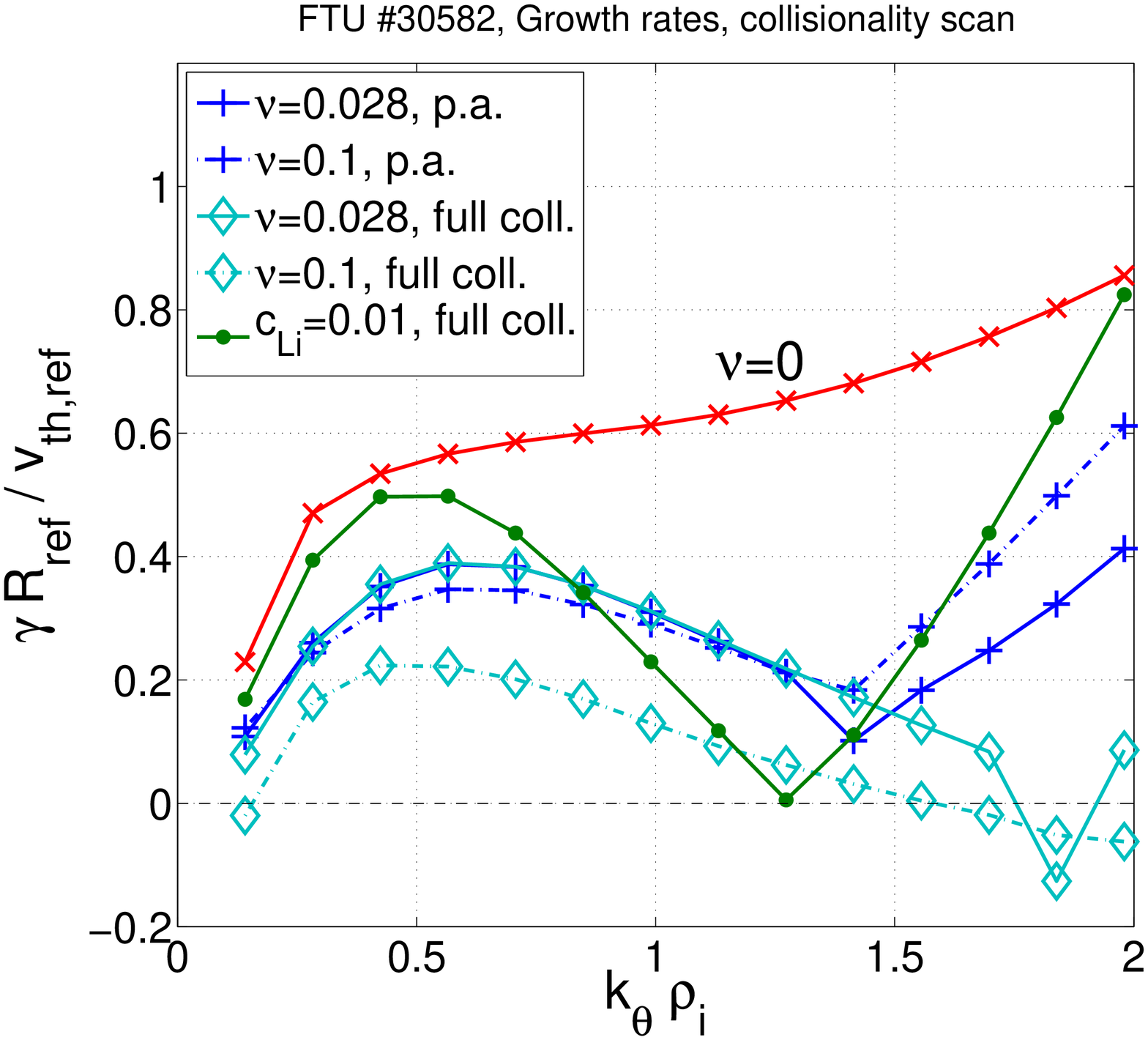}
  \includegraphics[scale=0.25]{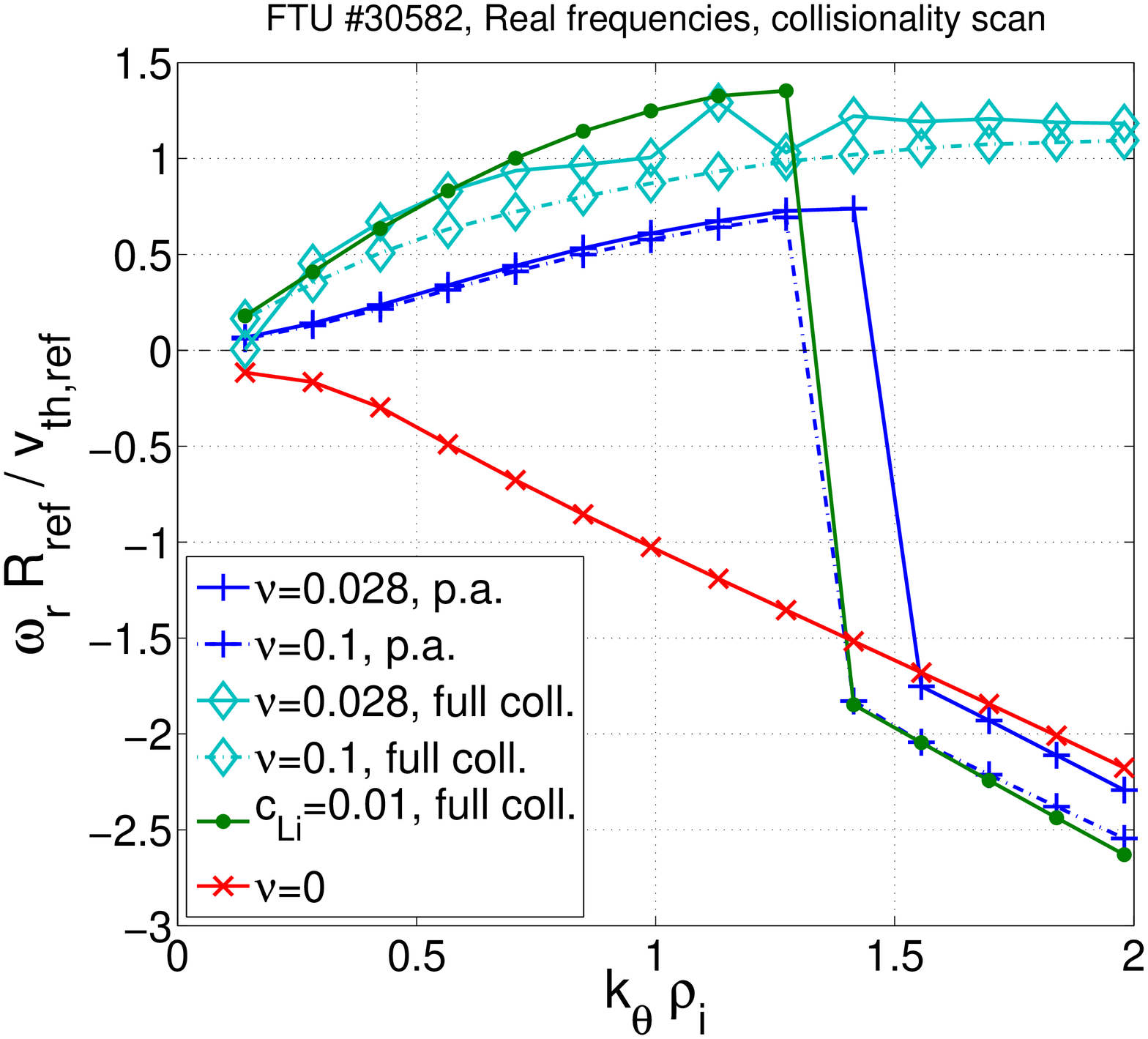} 
  \includegraphics[scale=0.25]{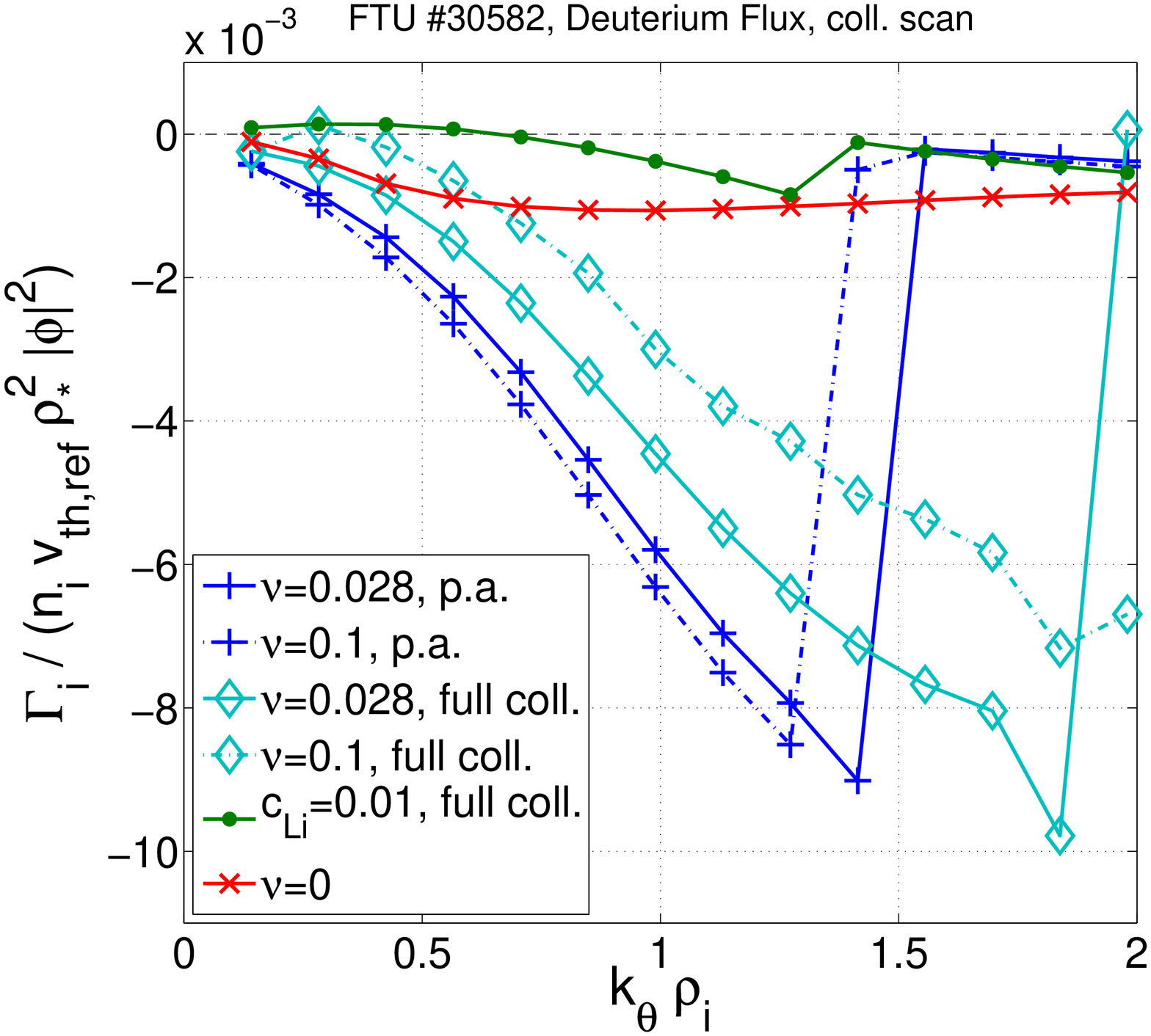}
  \includegraphics[scale=0.25]{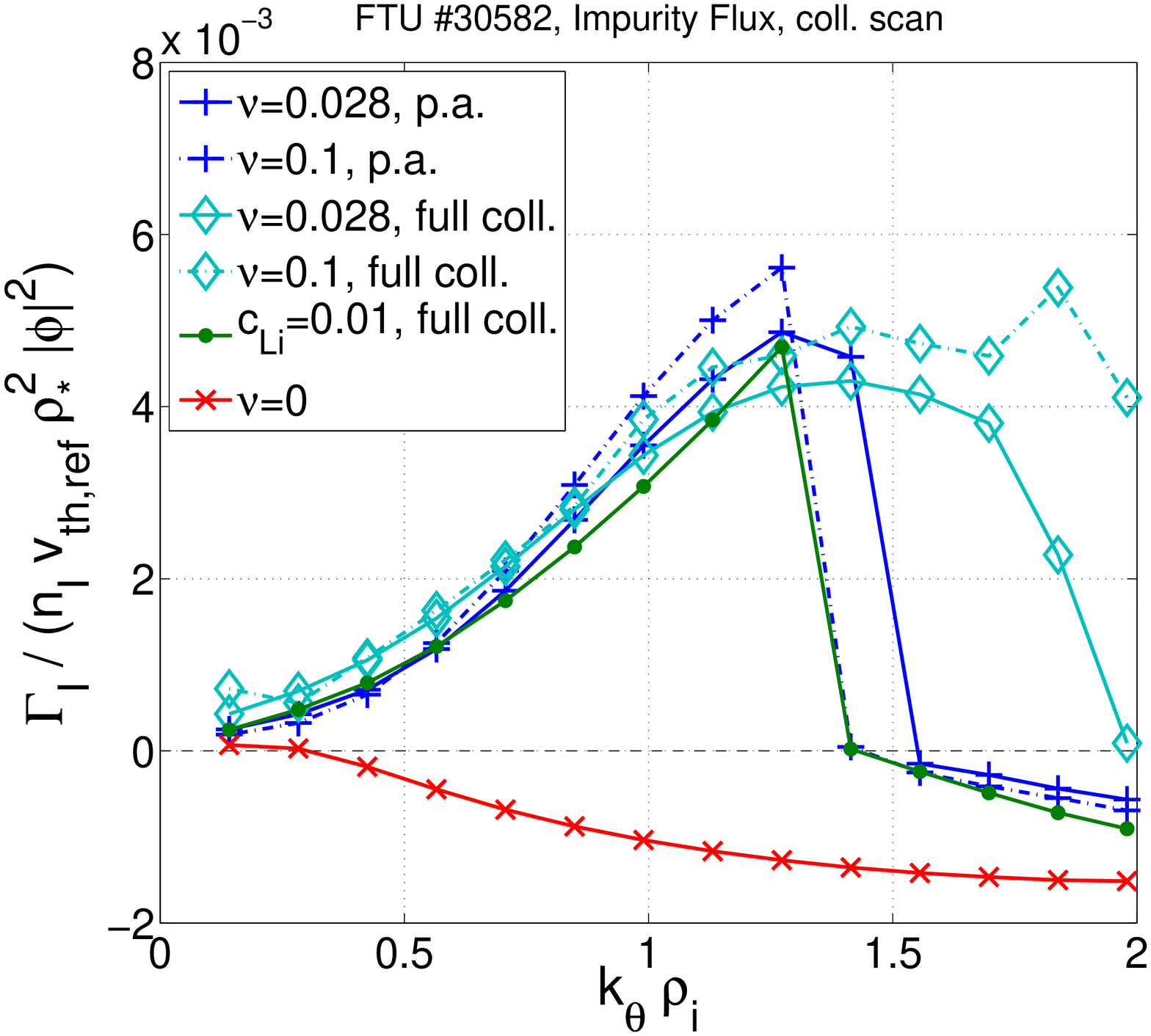}
 \end{center}
 \caption{Collisionality scan of growth rate (top left), real frequency (top right), quasi-linear deuterium (bottom left) and lithium (bottom right) flux as a function of bi-normal wavenumber at $t=0.3 \st{s}$. The reference impurity concentration is $c_{\st{Li}}=0.15$. Cases at the reference ($\nu_{\st{ii,N}}=0.028$, solid) and an increased ($\nu_{\st{ii,N}}=0.1$, dot dashed) collision frequency with both a Lorentz type (blue crosses) and the full (cyan diamonds) collision operator, together with a reduced impurity concentration ($c_{\st{Li}}=0.01$, solid, green dots) and a collisionless (red x-es) case are shown.}
 \label{fig:gk_lin_t0.3_nu}
\end{figure}

The top left and right panels of figure \ref{fig:gk_lin_t0.3_nu} show the linear growth rate and real frequency spectra of the bi-normal modes during the density ramp-up phase of the discharge ($t=0.3\st{s}$) for two different values of collision frequencies. Simulations performed with a Lorentz-type collision operator (pitch angle scattering only, crosses) and the full collision operator (pitch-angle scattering, energy scattering and friction terms, diamonds) are included. The collision frequencies are given in terms of the normalized ion-ion collision frequency $\nu_{\st{ii,N}}$ in units of $v_{\st{th,ref}}/R_{\st{ref}}$. The collision frequencies between the other species are calculated from $\nu_{\st{ii,N}}$ within the code \cite{gkw}. The reference value, as determined from the measured plasma parameters, is $\nu_{\st{ii,N}}=0.028$ (solid), and the increased value is $\nu_{\st{ii,N}}=0.1$ (dot dashed). Positive real frequencies correspond to modes rotating in the ion diamagnetic direction.
If collisions are not included in the simulation (red x-es), the spectrum is fully dominated by trapped electron modes. When using a Lorentz-type collision operator (blue crosses), the spectrum contains both ITG (below $k_{\theta} \rho_{\st{i}} \approx 1.4$) and TE modes. 
Increasing collisionality with respect to the reference value (dashed, crosses) weakly stabilizes ITG modes and destabilizes trapped electron modes. The latter observation was unexpected since TEM-s are typically stabilized by increasing collisionality even with pitch-angle scattering only. A similar behaviour is also observed at 1\% lithium concentration (not shown). A collisionality scaling of linear growth rates exhibiting a maximum has been reported in \cite{romanelli_2007_pop}, but there the increasing part of the $\gamma(\nu)$ function corresponds to significantly higher density gradients than those in the present simulation. 
When using the full collision operator (cyan diamonds), it changes the ITG/TEM spectrum to an ITG only case via stabilization of the TE modes. If the collision frequency is increased to $\nu_{\st{ii,N}}=0.1$ (dashed, diamonds), a uniform stabilizing effect across the whole spectrum is observed. If the collision frequency is further increased to the value of the $t=0.8 \st{s}$ case ($\nu_{\st{ii,N}}=0.22$) the modes are completely stabilized (not shown).  
When the impurity concentration is reduced to $c_{\st{Li}}=0.01$ while keeping the full collision operator (green dots), the reference collision frequency slightly changes through the weak density dependence of the Coulomb-logarithm (the change is less than 2\%). However, the frequency of the collisions with the two ion species is strongly affected: $\nu_{\st{si}}$ increases while $\nu_{\st{sI}}$ decreases due to the higher deuterium and lower impurity densities. The effective collision frequency of the species is proportional to the $Z_{\st{eff}}$ of the plasma. The reduced $Z_{\st{eff}}$ means that stabilization through collisions becomes weaker, the deuterium population less diluted, and the screening of the potential fluctuations less effective \cite{reshko}. These effects together cause the main ion ITG modes to grow more unstable, the impurity ITG driven tail to disappear and the electron driven modes to become dominant above $k_{\theta} \rho_{\st{i}} \approx 1.4$ (as observed in \cite{ftu_lin} with Lorentz collision operator).  

The corresponding quasi-linear deuterium and impurity flux spectra are shown on the bottom left and right panels of figure \ref{fig:gk_lin_t0.3_nu}, respectively. 
The flux is calculated as $\Gamma = \langle n_1 v_1 \rangle$ where $n_1$ and $v_1$ are the density and velocity fluctuations, and the angled brackets denote flux surface averaging. In linear simulations the potential is normalized with the mode amplitude squared at every timestep (included in the unit of the quasi-linear fluxes), and therefore the fluxes do not carry any information about their saturated values. The fluxes are proportional to the phase difference between the density and potential fluctuations during the linear phase of the mode evolution, and their signs indicate whether the flux is radially inward (negative) or outward (positive).
In the present case, the ITG modes are found to drive an inward deuterium and outward lithium flux at the reference impurity concentration $c_{\st{Li}}=0.15$, both with pitch-angle scattering and the full collision operator, at both collision frequencies. The direction of the particle transport remains the same even if the collision frequency is reduced to about one third of the reference value to $\nu_{\st{ii,N}}=0.01$ (not shown). However, at the reduced impurity concentration of $c_{\st{Li}}=0.01$ the modes below $k_{\theta} \rho_{\st{i}} \approx 0.6$ drive the deuterium ions outward. These modes are expected to provide the majority of the particle transport according to mixing length estimates and, consequently, to reverse the direction of the deuterium flux. This result is only obtained when the energy scattering and friction terms are taken into account in the collision operator and therefore it was missed in \cite{ftu_lin}. Although the impact of these terms on the linear fluxes are typically much smaller than that of the pitch-angle scattering operator, in this case it is enough to change the sign of the fluxes that are close to zero. The gyrokinetic simulations displayed in the remainder of the paper are performed using the full collision operator unless otherwise stated. 

\begin{figure}
 \begin{center}
  \includegraphics[scale=0.2]{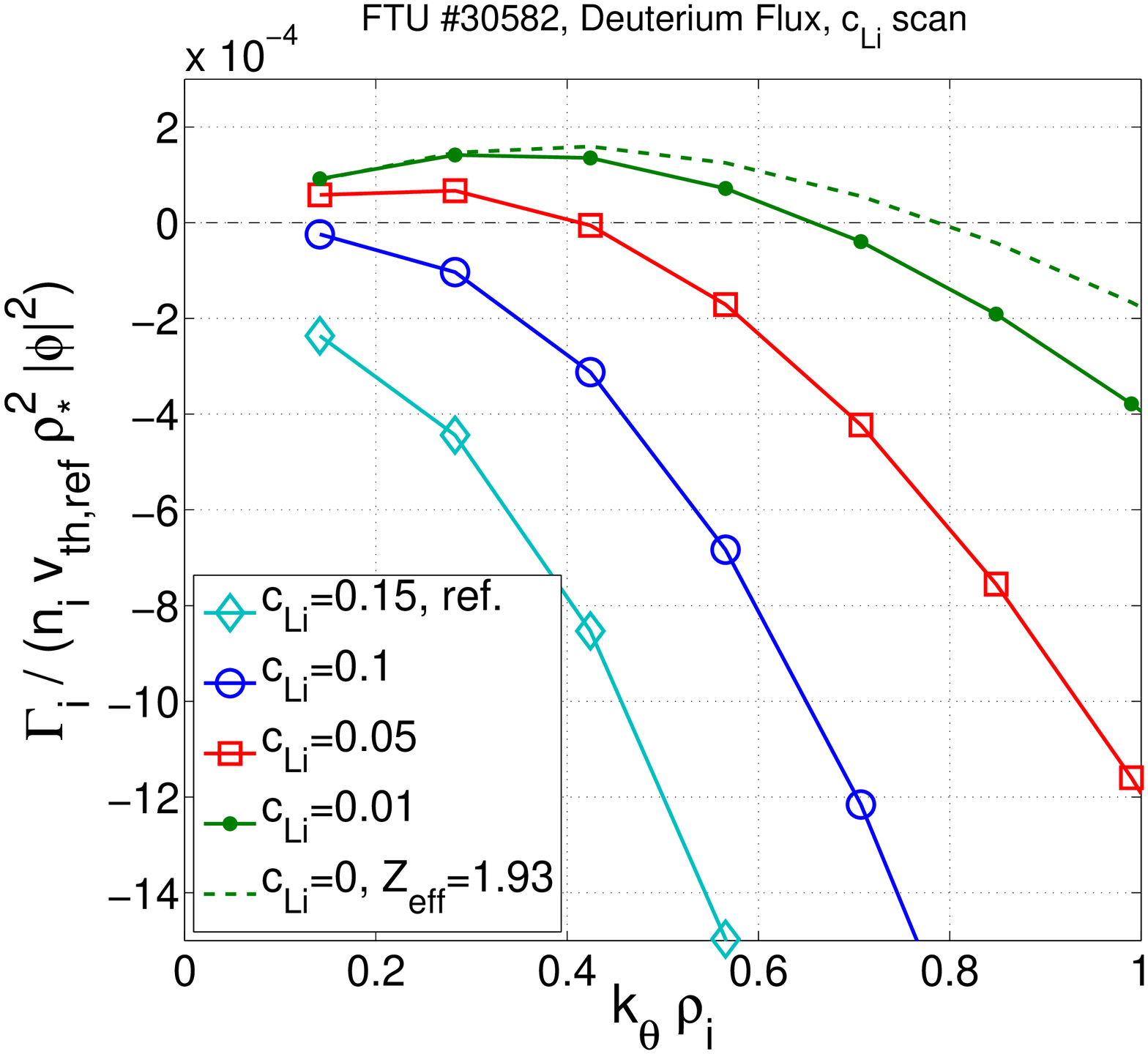}
  \includegraphics[scale=0.2]{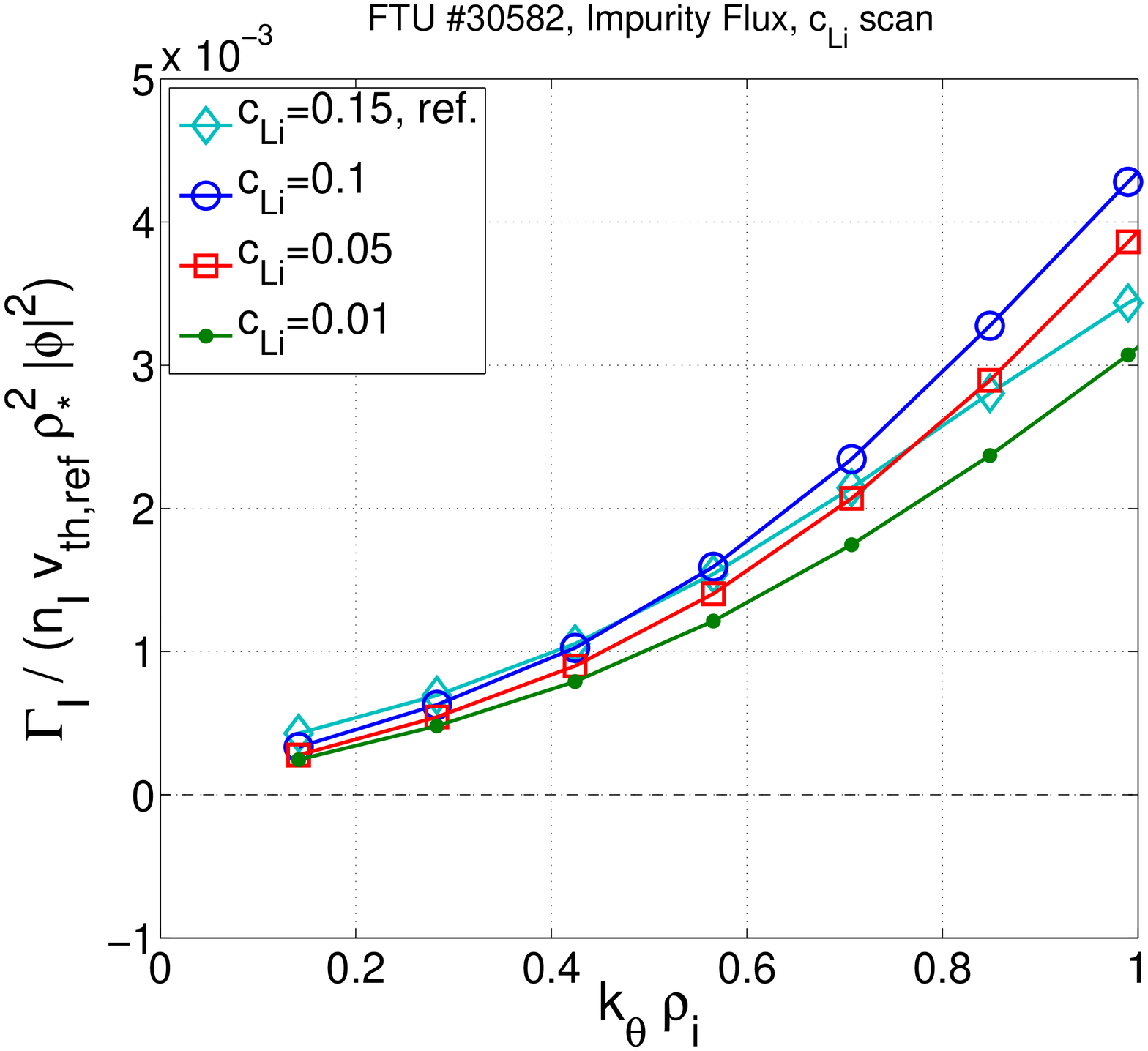}
  \includegraphics[scale=0.2]{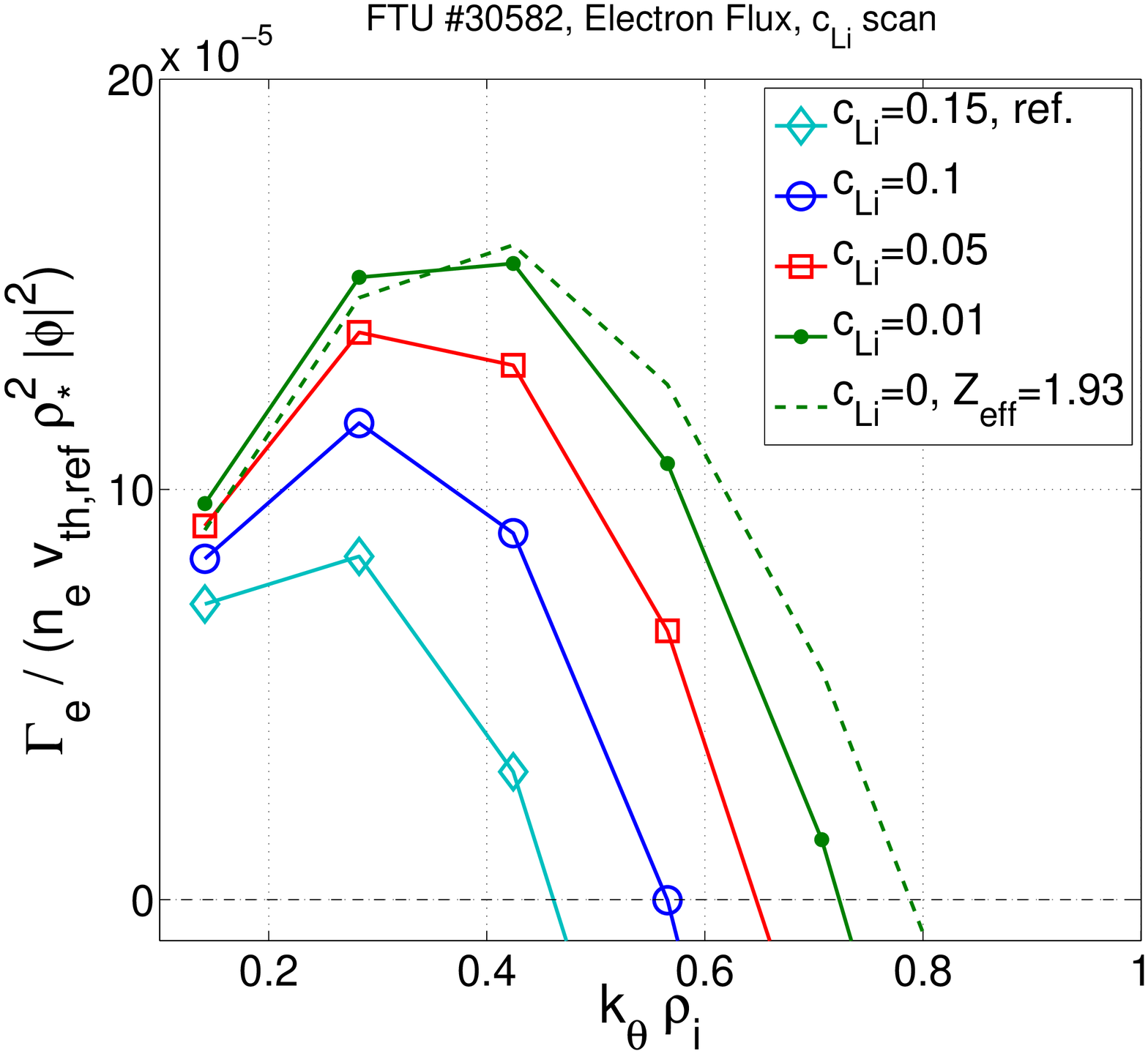}
 \end{center}
 \caption{Lithium concentration scan of quasi-linear deuterium (left), lithium (middle) and electron (right) quasi-linear flux spectra as a function of the bi-normal wavenumber at $t=0.3 \st{s}$.}
 \label{fig:gk_lin_t0.3_nLi}
\end{figure}

A scan of the impurity concentration in the $t=0.3 \st{s}$ case is shown on figure \ref{fig:gk_lin_t0.3_nLi}. The trend of the growth rates (not shown) is similar to as described in \cite{ftu_lin} with pitch-angle scattering only. 
However, below a certain lithium concentration the low-k modes (below $k_{\theta} \rho_{\st{i}} \approx 0.5$ at $c_{\st{Li}}=0.1$) drive an outward deuterium flux (left). Inward deuterium transport requires approximately $c_{\st{Li}} > 0.05$ in this case. The outward impurity flux driven by the ITG modes (middle) remains unchanged. As the impurity concentration is increased the outward electron flux driven by the low-k modes (right) is also reduced. This can be attributed to the fact that the electrons are less sensitive to the ion scale modes, and tend to remain nearly adiabatic in ITG turbulence. When the impurity population is low the electrons must follow the ion transport due to quasi-neutrality. However, when the quasi-neutrality equation contains three major constituents, the restriction imposed by ambi-polarity can be satisfied with the two ion species.
The dashed (green) curve on the left and right panels show a pure plasma simulation with zero impurity concentration, but with $Z_{\st{eff}}$ manually set to its original value of 1.93. This provides approximately the same effective electron and ion collision frequencies as in the experimental case (within 2\% difference due to the change of the Coulomb logarithm). The growth rates, frequencies and fluxes are nearly identical to the 1\% impurity concentration results, showing that the main impact of the lithium is not through collisional effects. 
 
\begin{figure}
 \begin{center}
  \includegraphics[scale=0.25]{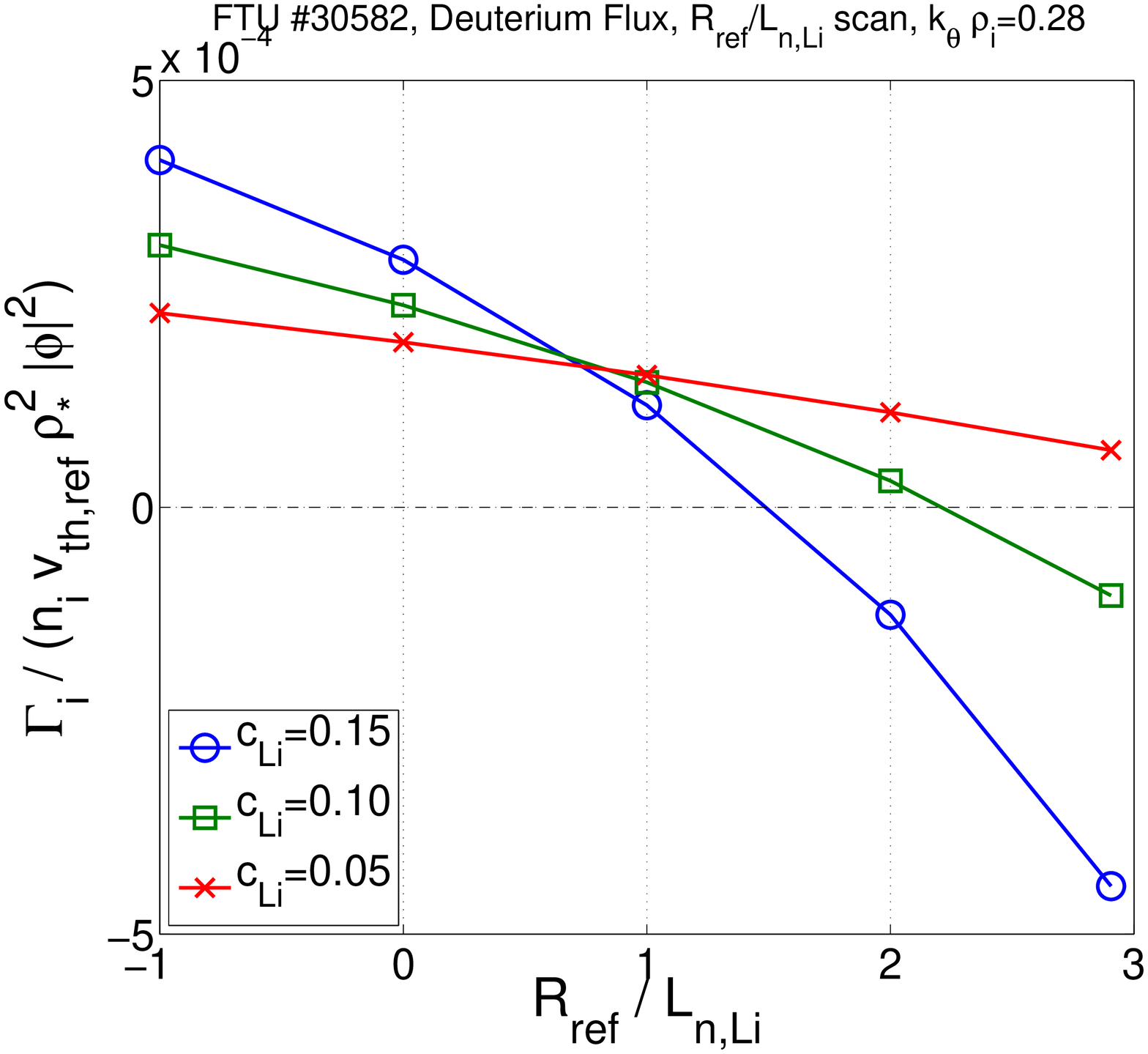}
  \includegraphics[scale=0.25]{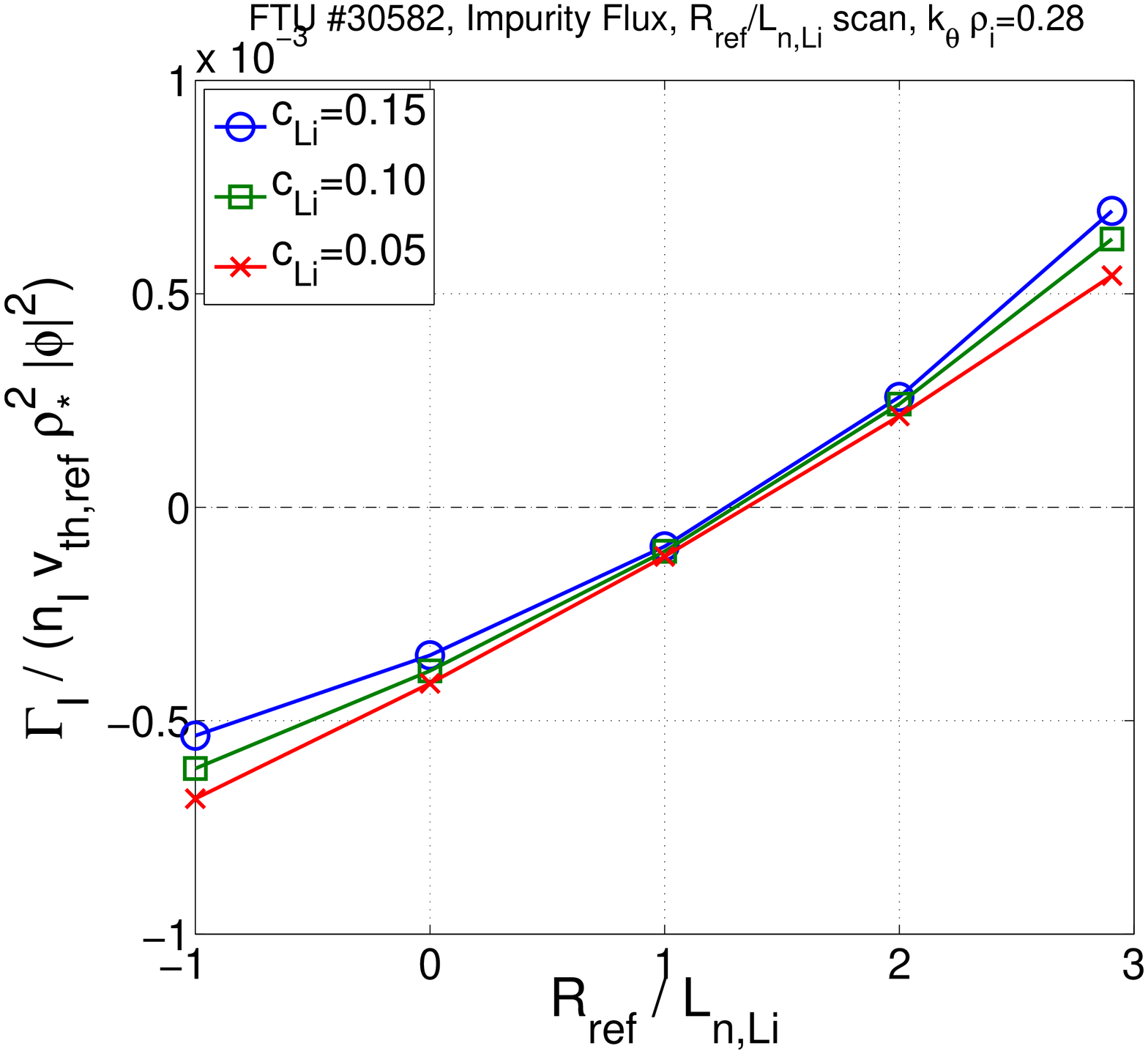}
 \end{center}
 \caption{Impurity concentration and impurity density gradient scan of the quasi-linear deuterium (left) and lithium (right) fluxes at $t=0.3 \st{s}$ and $k_{\theta} \rho_{\st{i}}=0.28$.}
 \label{fig:gk_lin_t0.3_RLnLi_nLi_scan}
\end{figure}

It has been pointed out by several authors (e.g. \cite{emila,frojdh,coppi1976}) that impurity transport is sensitive to the ion density gradient. 
In the previous simulations the same positive density gradient value has been used for all three species, that is a centrally peaked impurity density profile has been assumed. However, experimentally the impurity profile is often found to be peaked towards the edge. 
The effect of varying the lithium density gradient on the quasi-linear ion flux at three different impurity concentrations is investigated in figure \ref{fig:gk_lin_t0.3_RLnLi_nLi_scan}. Since the flux reversal has been previously observed at low-k modes (typically below $k_{\theta} \rho_{\st{i}} \approx 0.5$), and these are the modes that provide the largest contribution to the saturated flux, the $k_{\theta} \rho_{\st{i}} = 0.28$ mode has been selected for this analysis. 
Since the electrons are expected to be close to adiabatic in a highly contaminated plasma, the direction of the deuterium and lithium transport is determined by the balance of the drive of their respective fluxes. This drive strongly depends on the concentration and the density gradient of the impurity species, while the impact of these parameters on the low-k ITG mode stability is relatively weak. 
If the impurity concentration is sufficiently high and its density profile sufficiently peaked, such as it is assumed during the density ramp-up of this FTU-LLL discharge, then the outward drive of the impurity flux outweighs that of the deuterium, and forces the main ions to be transported inward. This observation is qualitatively consistent with the results of \cite{emila} obtained with helium impurities. 

\begin{figure}
 \begin{center}
  \includegraphics[scale=0.25]{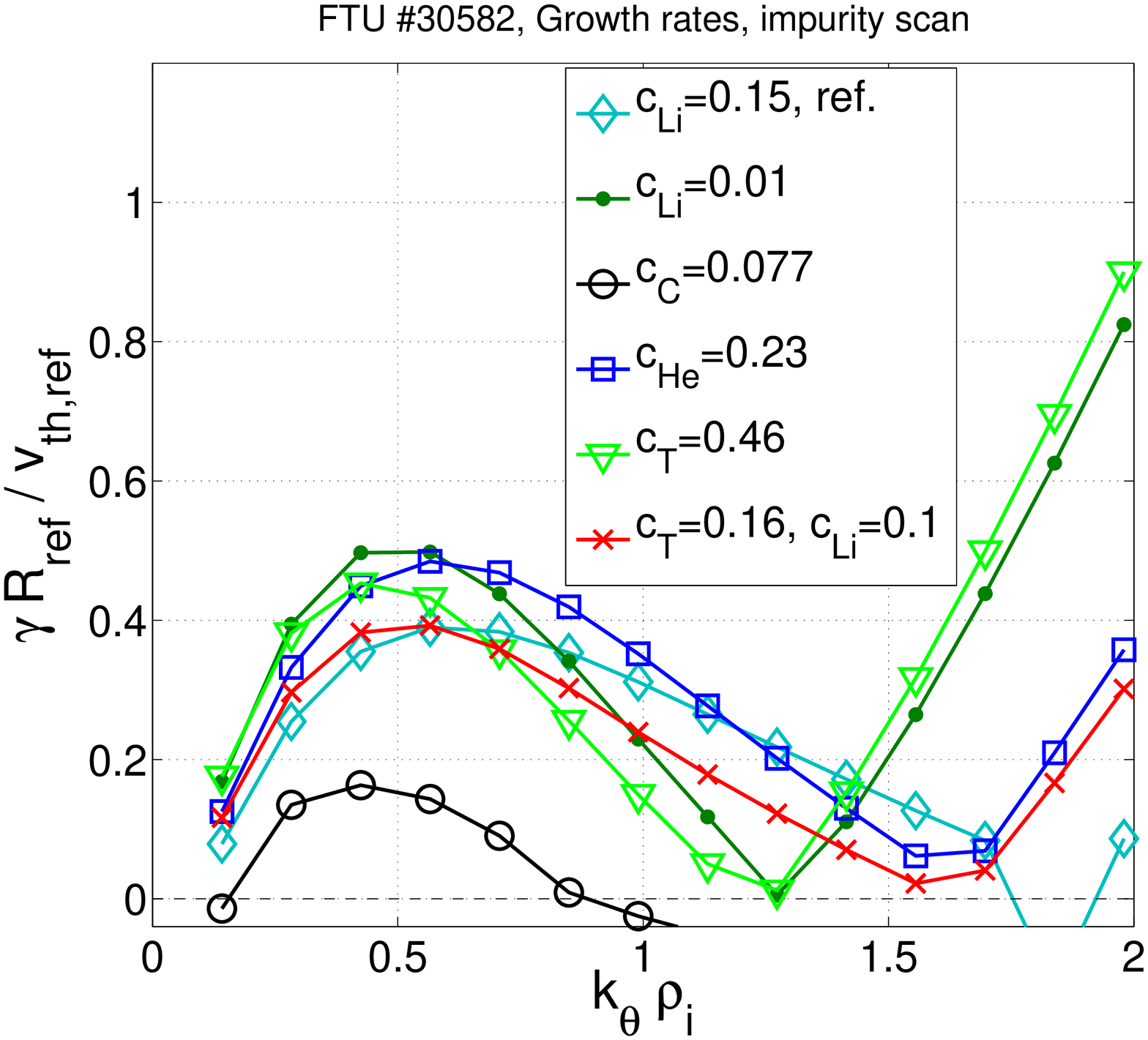}
  \includegraphics[scale=0.25]{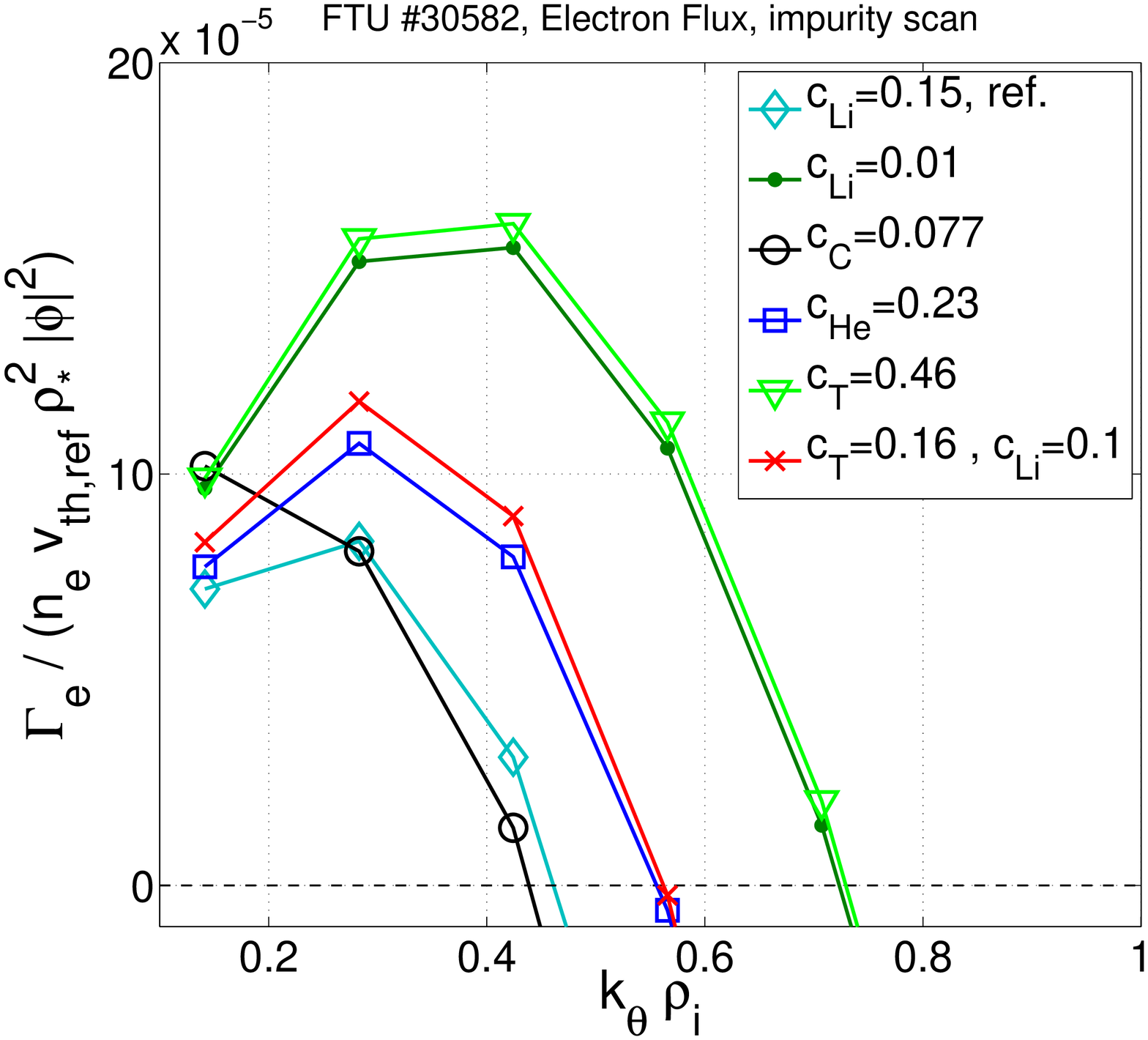} 
  \includegraphics[scale=0.25]{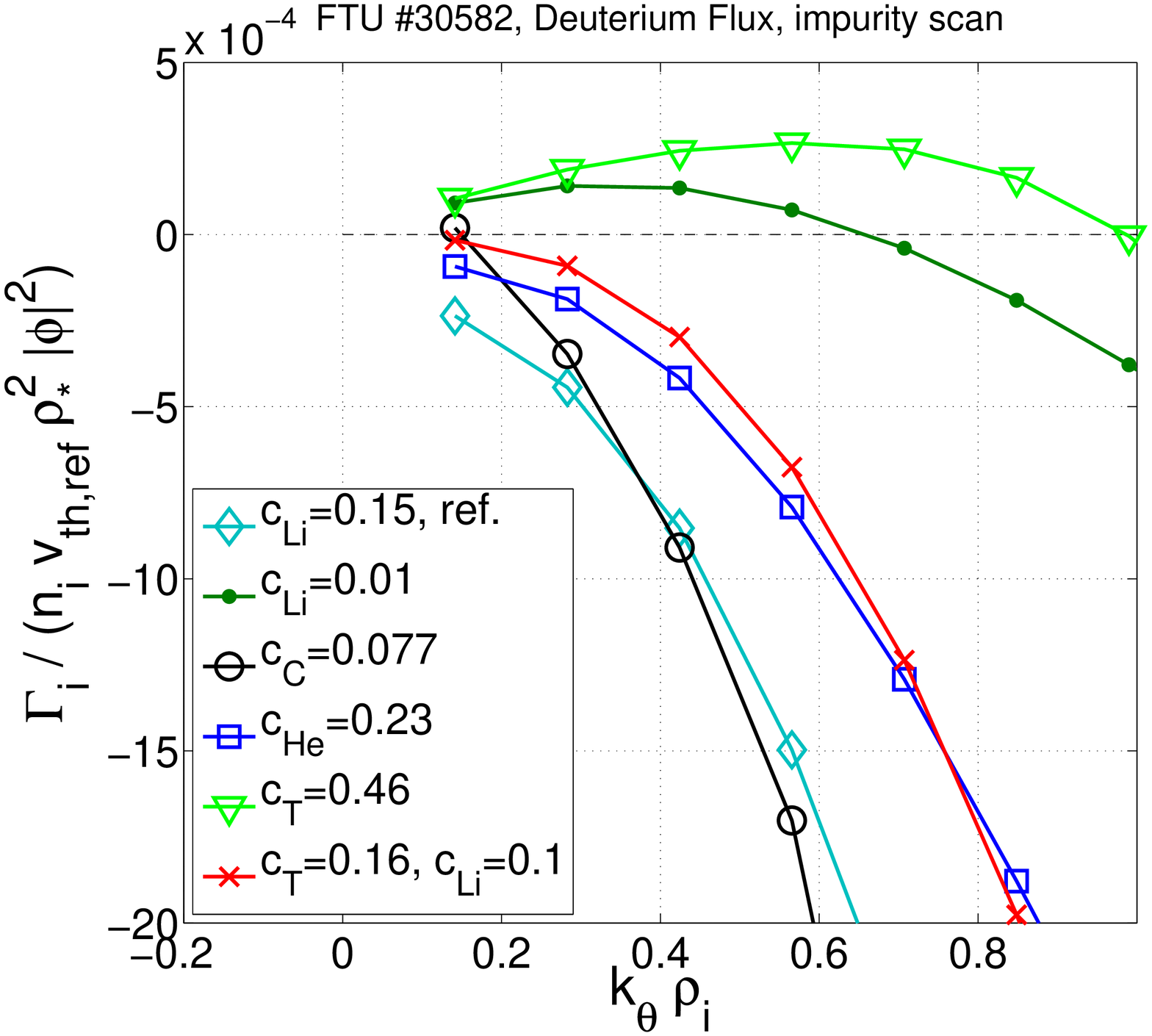}
  \includegraphics[scale=0.25]{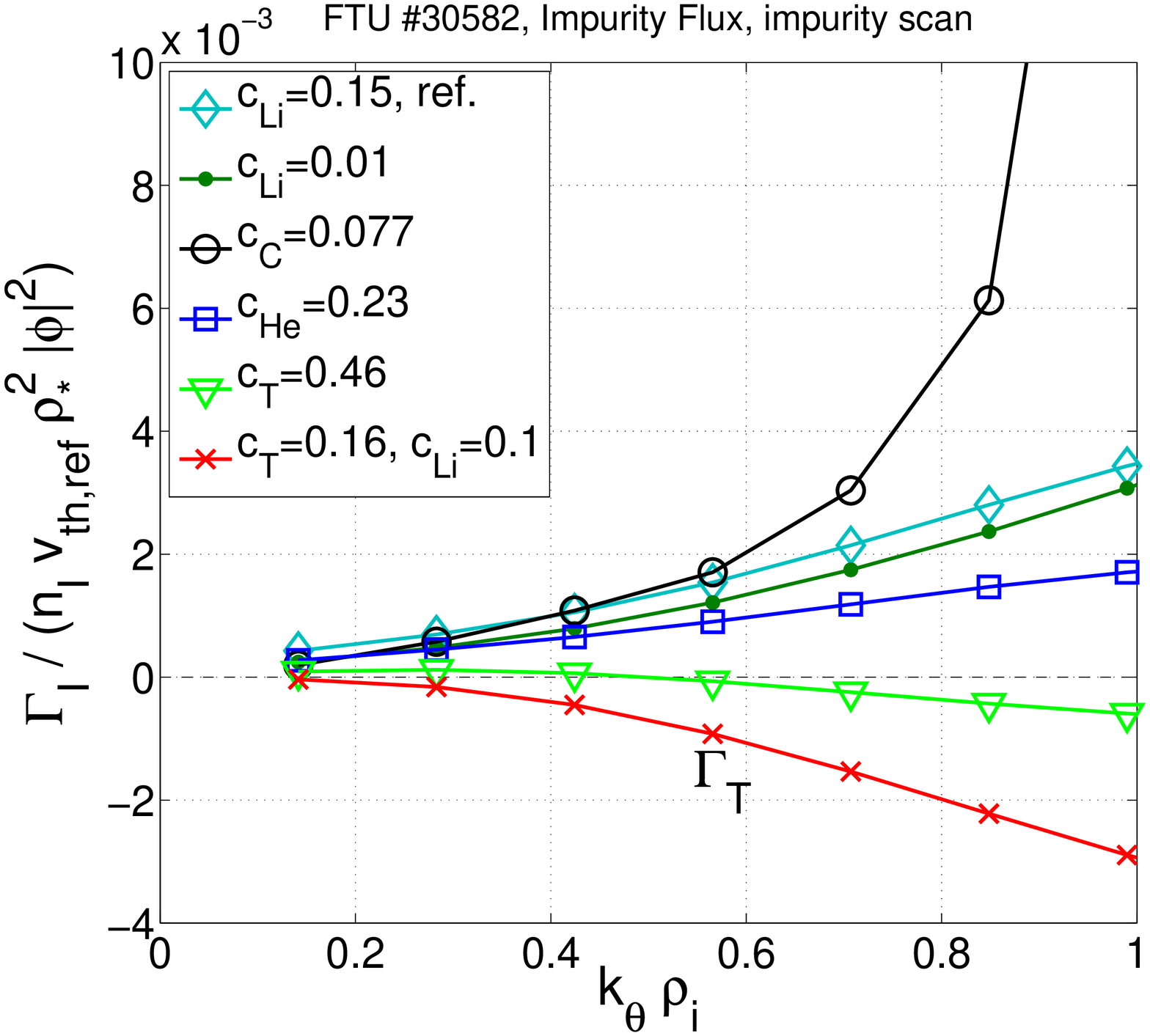}
 \end{center}
 \caption{Growth rate (top left), electron (top right), deuterium (bottom left) and impurity (bottom left) quasi-linear flux spectra as a function of the bi-normal wavenumber using different impurity species.}
 \label{fig:gk_lin_t0.3_impurity_scan}
\end{figure}

The significance of the impurities being lithium ions is tested by a series of simulations using different light impurity species (carbon, tritium, helium) while keeping the deuterium concentration constant. A mixed deuterium-tritium-lithium case is also investigated due to its potential relevance in future burning plasma applications. The results are plotted in figure \ref{fig:gk_lin_t0.3_impurity_scan}. 
The carbon (black circles) and lithium (cyan diamonds) are the most effective among the selected impurity ions in reducing the quasi-linear electron flux and driving the deuterium flux inward. However, under the current circumstances, the increase of the ion collisionality due to 7\% carbon concentration strongly stabilizes the modes. The helium impurity (blue squares) and the mixed D-T-Li case exhibit similar behaviour, but the results with 46\% tritium (light green triangles) are close to the clear plasma simulations (1\% lithium, green dots), both of them driving an outward deuterium flux. 
If the deuterium and impurity Larmor-radii are too close, they are expected to react similarly to the main ion ITG modes, and thus to be unable to reduce the electron transport and generate the deuterium pinch. 
Indeed, the quasi-linear tritium flux in both the D-T and D-T-Li cases follow that of the deuterium, although in the former case the flux is close to zero and only the three lowest k modes indicate an outward transport. 
The dynamics of the deuterium and lithium ions are sufficiently different for the ion flow separation to occur. However, the typical D-ITG length scale is still sufficiently close to the lithium Larmor-radius that the lithium ions can have a significant effect on the low-k ITG transport. Moreover, $Z_{\st{Li}}$ is low enough that it does not have a major qualitative impact on mode stability and it does not cause substantial radiative losses. 

Increasing the magnetic shear in the range of $\hat{s}=0.5-1.5$ stabilizes the ITG modes, also observed in more detailed studies on the impact of magnetic shear by Kinsey et al. \cite{kinsey}, Moradi at al. \cite{moradi} and Nordman at al. \cite{nordman_imp}. However, we note that the actual effect of the magnetic shear strongly depends on the other plasma parameters and its stabilizing effect should be considered on a case-to-case basis. TEM-s are suppressed by the collisionality and they remain stable even at the lower shear value. The phase difference between potential and ion density perturbations depends weakly on the variation of the magnetic shear.

\subsubsection{The Density Plateau Phase}

At $t=0.3\st{s}$ the plasma is not in steady-state. It is characterized by finite radial particle transport that, as it has been shown above, contributes to the dynamical build-up of the deuterium density profile and the reduction of impurity concentration. At $t=0.8\st{s}$ the plasma is described by significantly larger deuterium and electron density and density gradients, reduced $Z_{\st{eff}}$, and a slightly increased magnetic shear. The reduced temperature is likely to be caused by the inward flux of cold deuterium ions from the edge. The reference collision frequency is approximately a factor of eight higher ($\nu_{\st{ii,N}}=0.22$) due to the lower temperature and higher density values, leading to an increased effective collision frequency of the species despite the lower $Z_{\st{eff}}$.
It was shown in \cite{ftu_lin} that increasing only the density gradient of the species from their values at $t=0.3 \st{s}$ to that at $t=0.8 \st{s}$ would be sufficient to obtain an outward flux of all three plasma constituents, as observed during the density plateau phase. However, at larger density gradients the so called ion mixing modes \cite{coppi1990} (characterized by monotonously increasing mode frequency starting at negative values at low $k_{\theta}$) become dominant across the whole spectrum, which is in contrast with the $t=0.8 \st{s}$ simulations. 

\begin{figure}
 \begin{center}
  \includegraphics[scale=0.2]{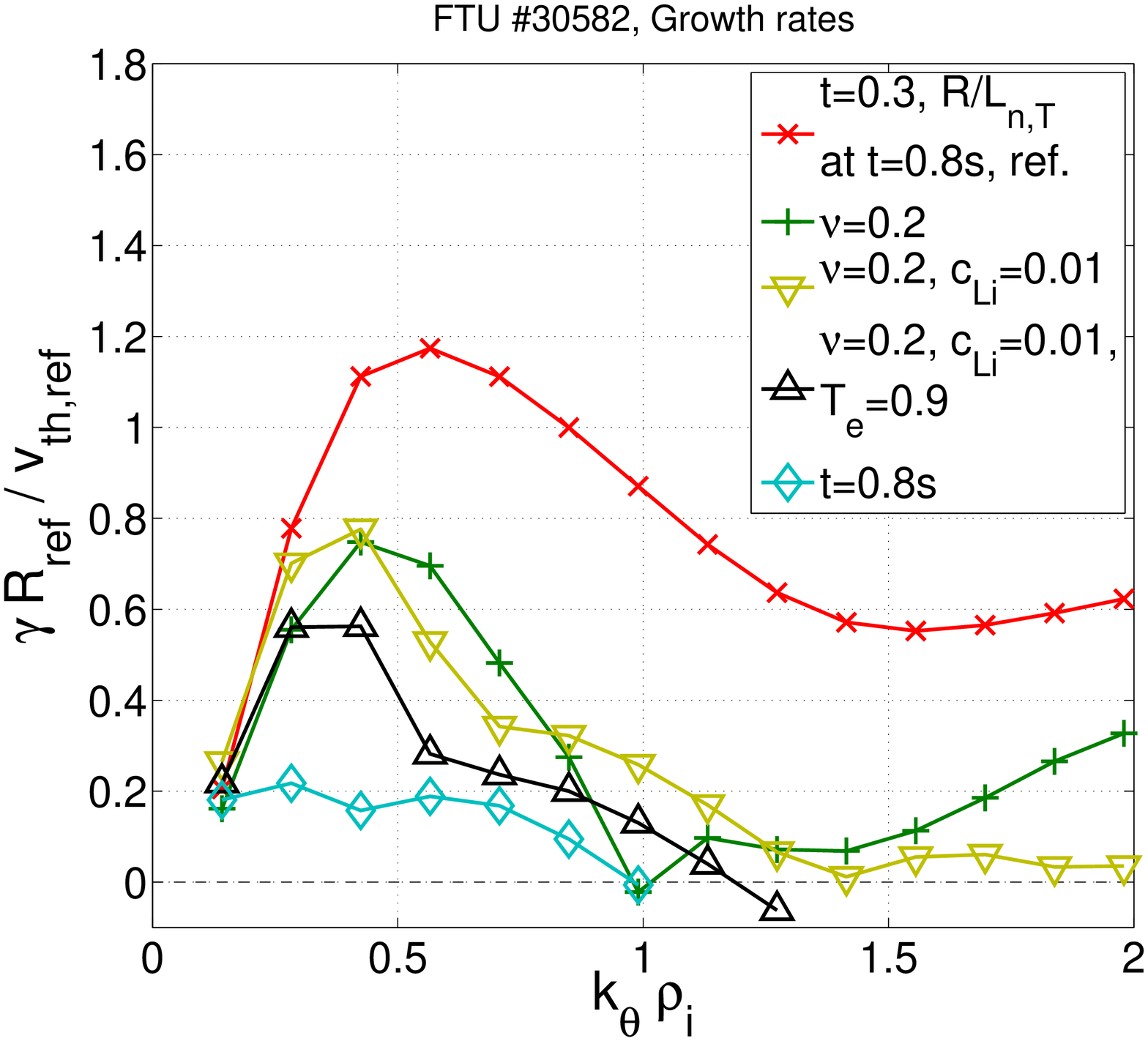}
  \includegraphics[scale=0.2]{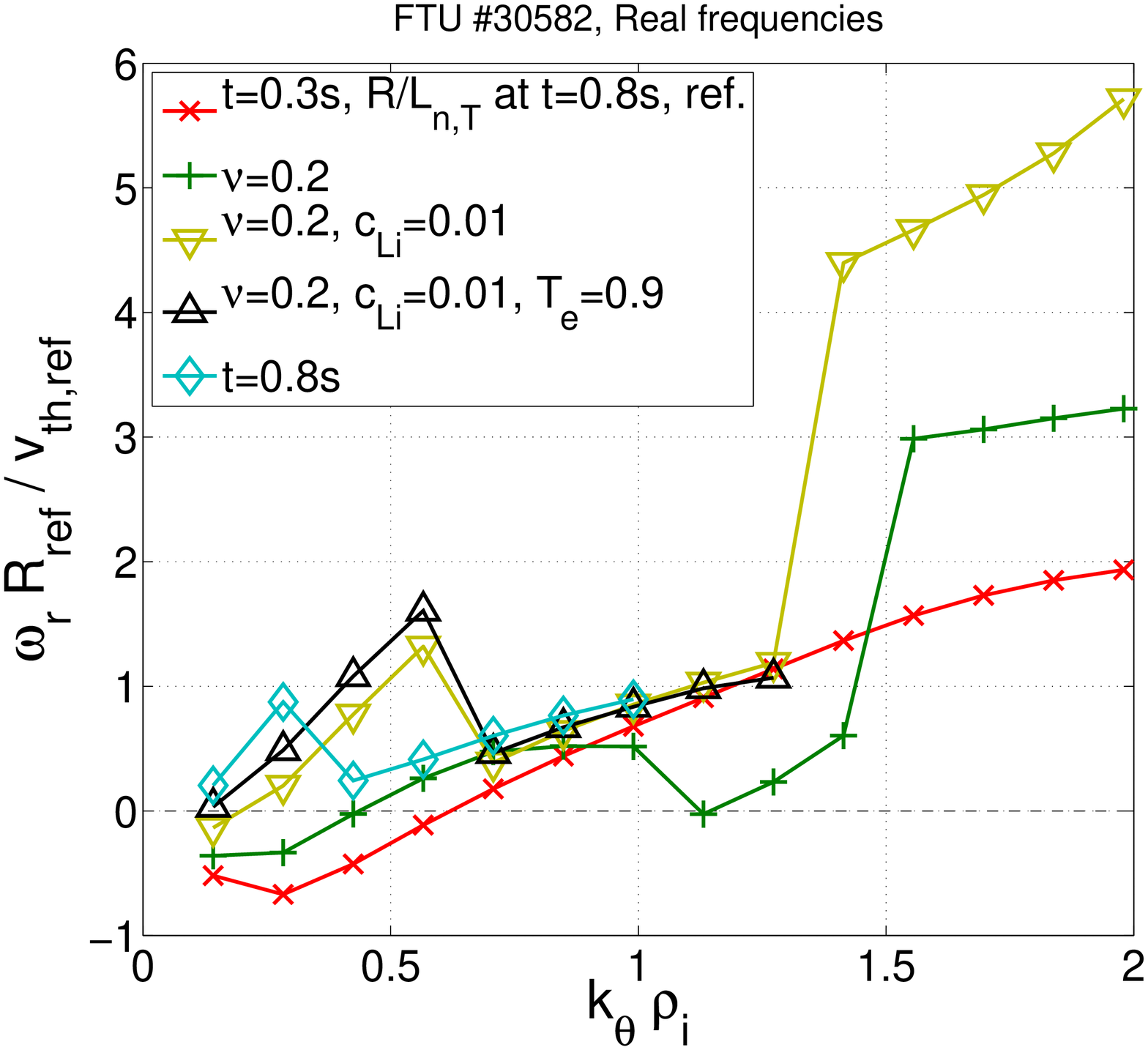} 
  \includegraphics[scale=0.2]{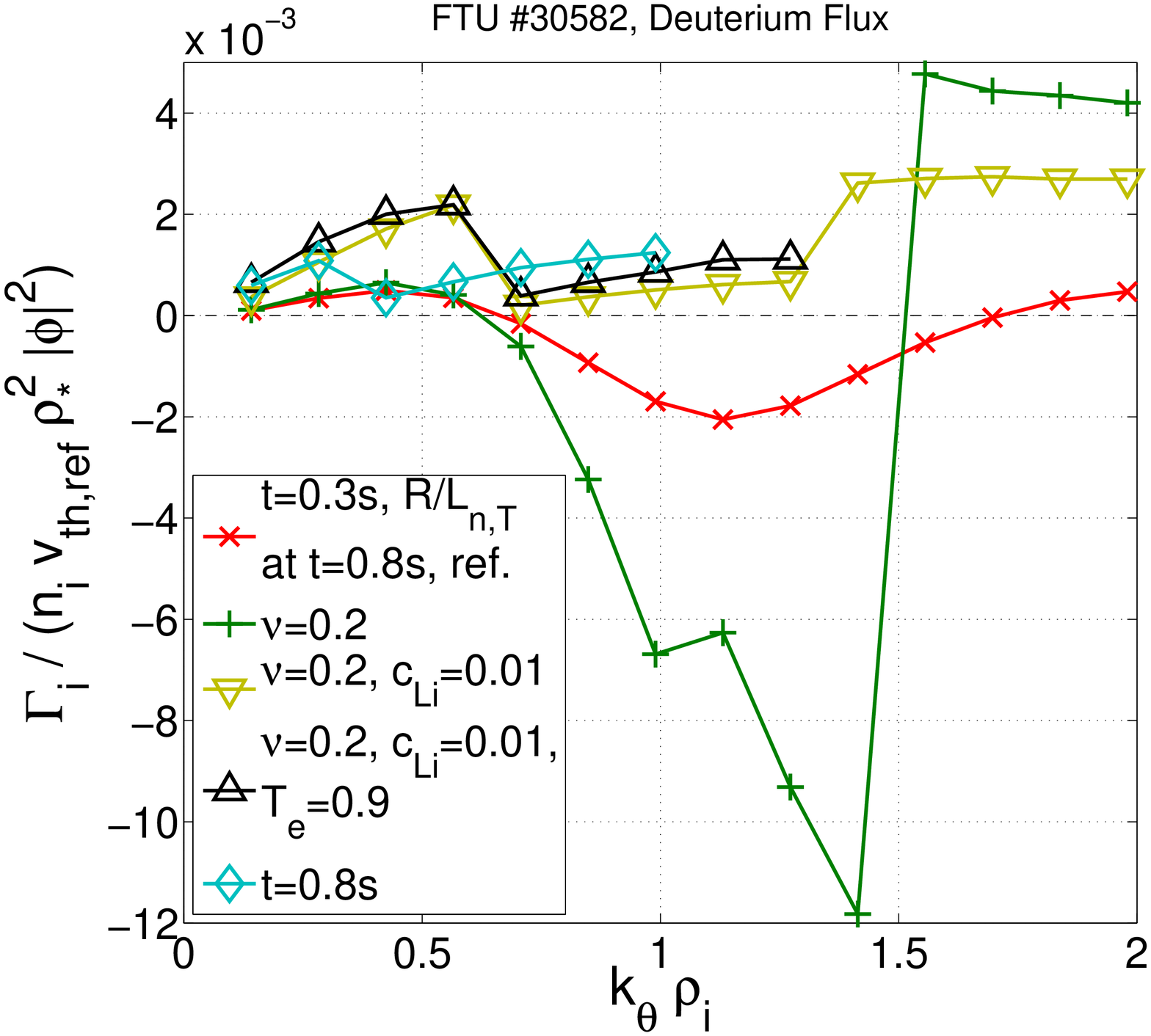}
 \end{center}
 \caption{Approaching the $t=0.8 \st{s}$ case (cyan diamonds) by increasing the gradients (red x-es), collisionality (green crosses), decreasing the lithium concentration (yellow down triangle) and electron temperature (black up triangle) from their respective values at $t=0.3 \st{s}$.}
 \label{fig:gk_lin_t0.3_to_t0.8}
\end{figure}

The microstability and particle transport properties of the $t=0.8 \st{s}$ case can be further approached by sequentially increasing the reference collision frequency to $\nu_{\st{ii,N}}=0.2$, decreasing the lithium concentration to $n_{\st{Li} / n_{\st{e}}} = 0.01$ and electron temperature to $T_{\st{e}} / T_{\st{i}} = 0.9$, as shown on figure \ref{fig:gk_lin_t0.3_to_t0.8}. The graphs indicate that the high collisionality is essential in stabilizing the ion mixing modes but it is not, on its own, sufficient to obtain the mode structure and fluxes as observed at the later time stage. Only by reducing the lithium concentration (and thus decreasing the effective collisionality) can an ITG dominated spectrum and outward deuterium flux in the mid-k range ($k_{\theta} \rho_{\st{i}} > 0.6$) be reproduced. The high-k modes are stabilized by the lower electron temperature, and the low-k ITG mode growth rates are reduced by the increased magnetic shear (not shown), finally providing the spectrum of the $t=0.8 \st{s}$ case.

\begin{figure}
 \begin{center}
  \includegraphics[scale=0.2]{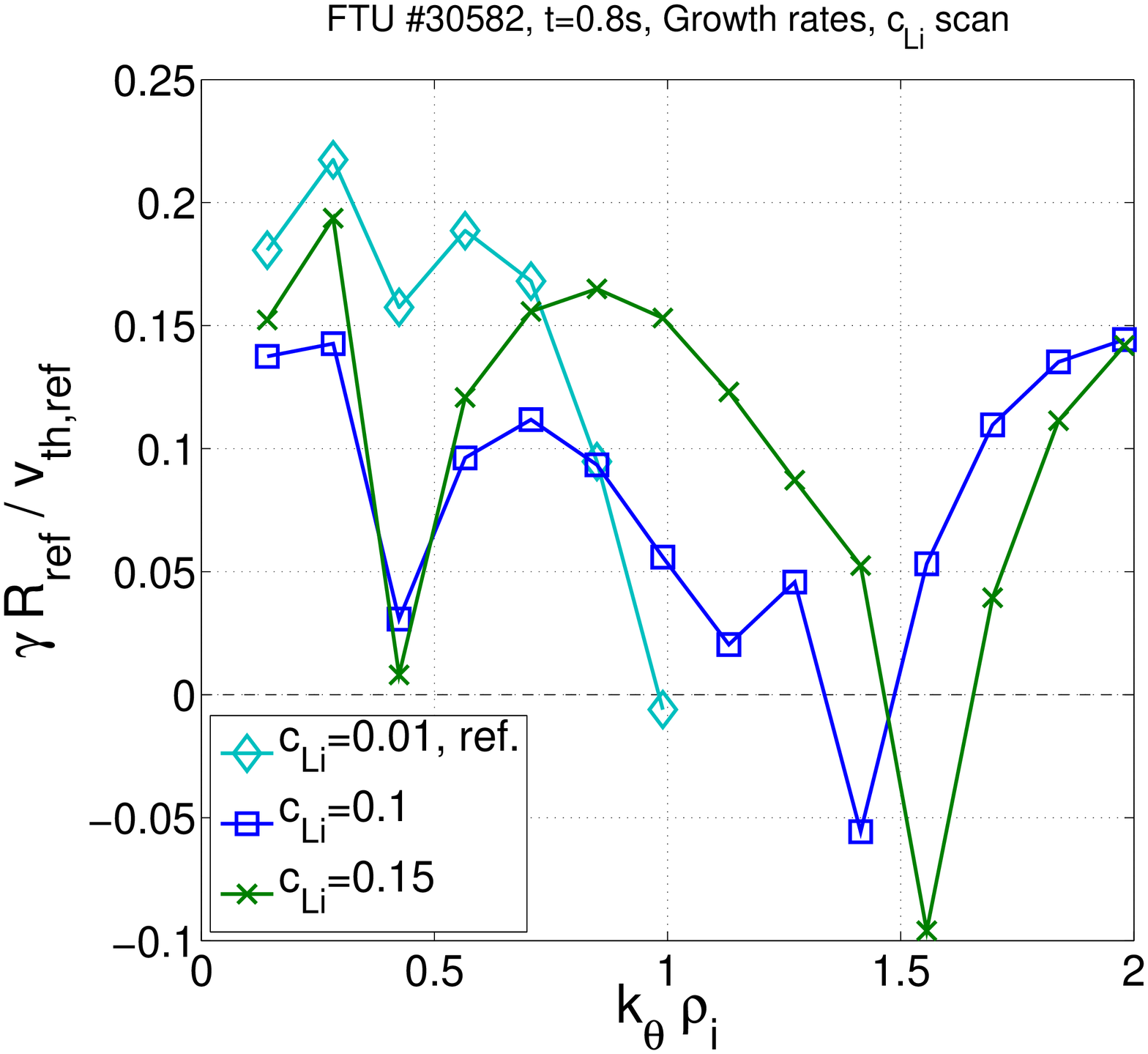}
  \includegraphics[scale=0.2]{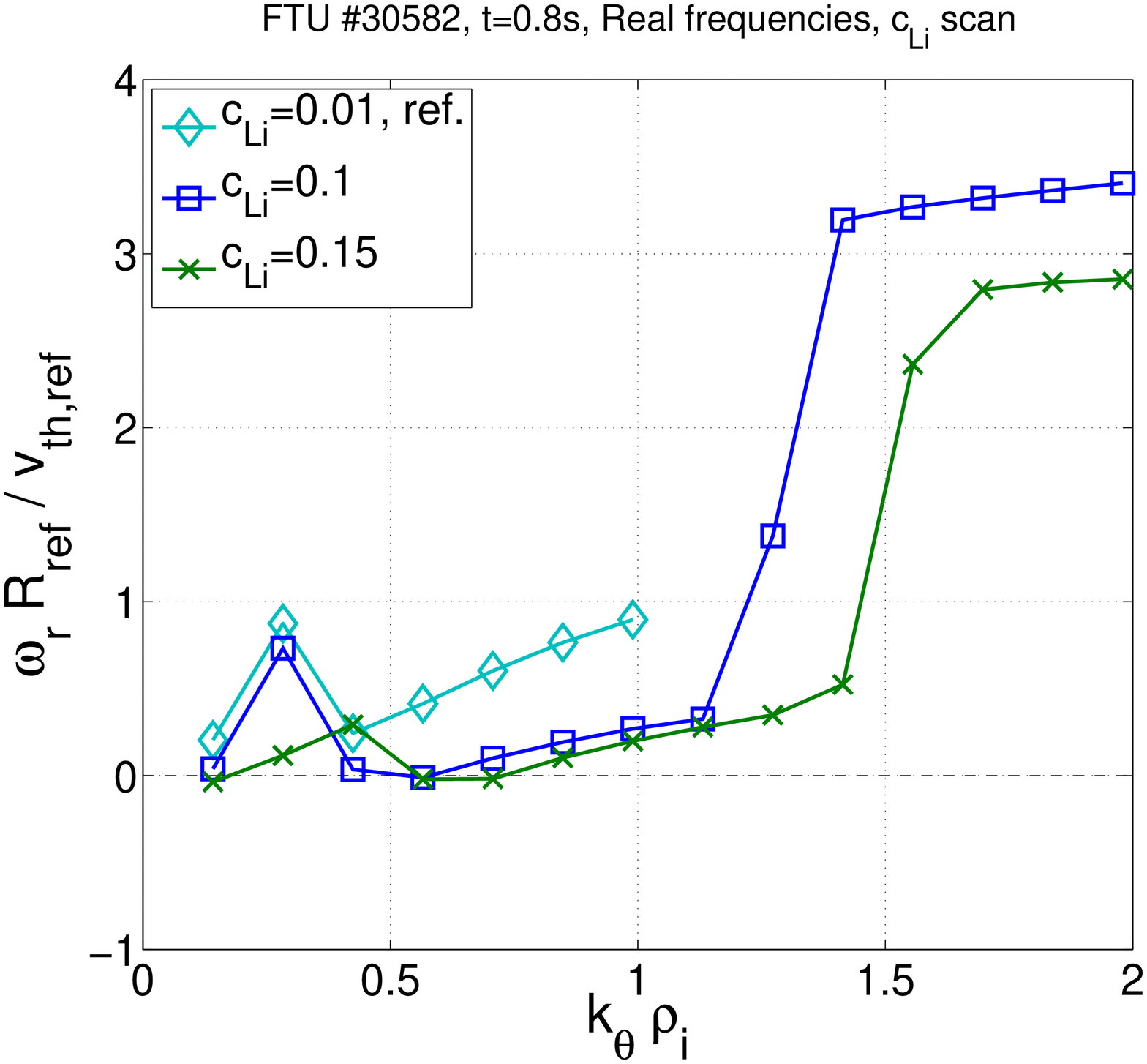} 
  \includegraphics[scale=0.2]{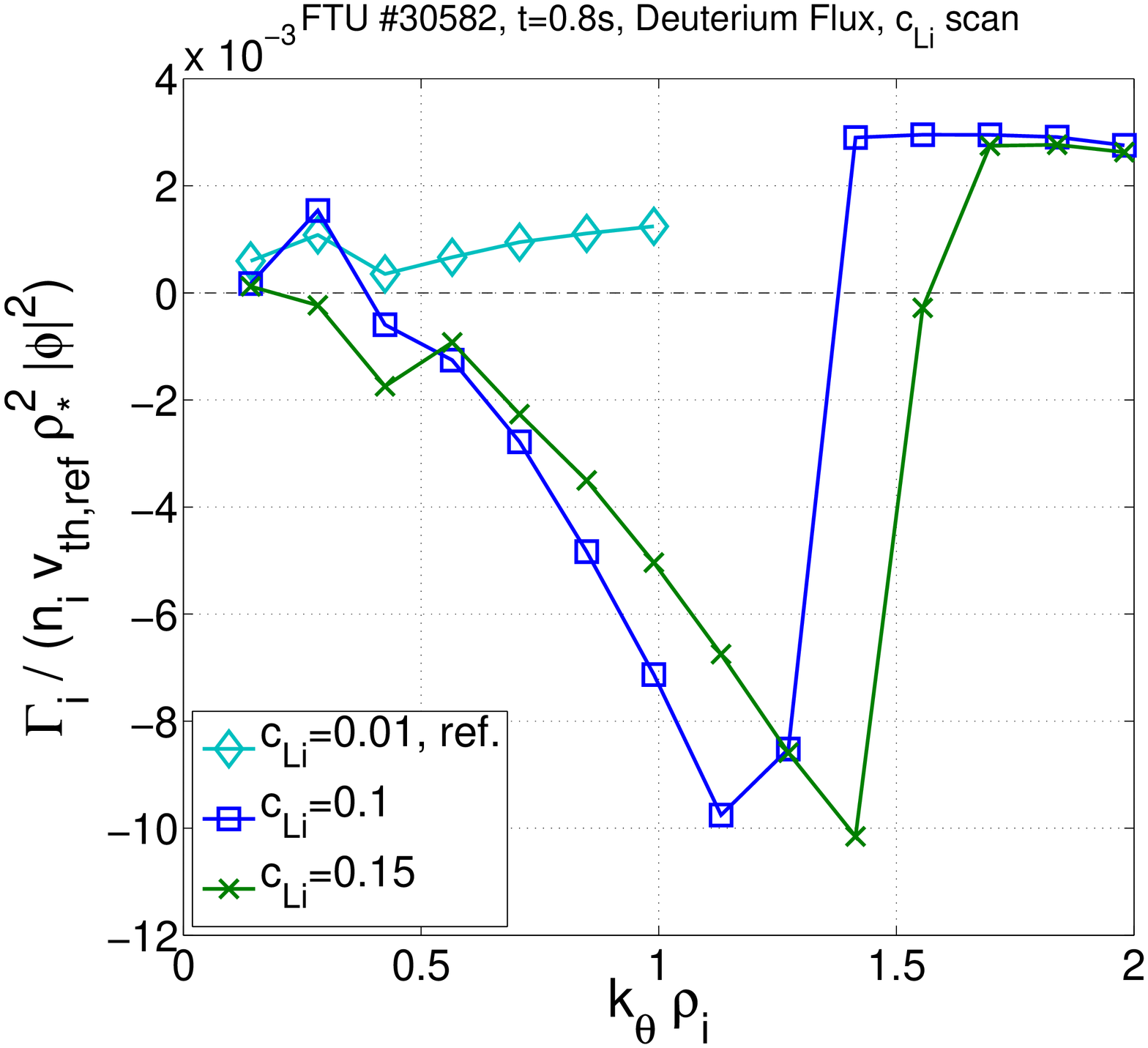}
 \end{center}
 \caption{Lithium concentration scan of growth rate (left), real frequency (middle) and quasi-linear deuterium flux (right) as a function of bi-normal wavenumber at $t=0.8 \st{s}$.}
 \label{fig:gk_lin_t0.8_nLi}
\end{figure}

The impurity density scan at $t=0.8 \st{s}$, presented on figure \ref{fig:gk_lin_t0.8_nLi}, shows that the reversal of the deuterium flux can be retained again with high lithium concentration. However, the effect on the microstability is even more pronounced in this case. 
Both at $c_{\st{Li}}=0.1$ and $c_{\st{Li}}=0.15$ the growth rate (left) and frequency (middle) spectra suggest the presence of three distinct modes rotating in the ion direction. 
The low-k peak can be associated with the D-ITG modes, the middle peak between $0.5<k_{\theta} \rho_{\st{i}}<1.5$ with Li-ITG modes, and above $k_{\theta} \rho_{\st{i}}\approx 1.5$ with ion drift modes. The stabilization of the D-ITG and destabilization of the ion drift modes is attributed to the significantly increased effective collisionality. 
The ion drift modes drive an outward deuterium flux but their contribution is likely to be small in the saturated phase due to their high $k_{\theta}$ values.  

\subsubsection{Discussion}

At $t=0.3 \st{s}$ D-ITG and Li-ITG modes are not as clearly separated in the bi-normal wavenumber spectrum (figure \ref{fig:gk_lin_t0.3_nu}), they are expected to be mixed ITG modes sensitive to the temperature gradients of both ion species \cite{ftu_lin}. However, the presence of the lithium ions changes the phase difference between the deuterium density and potential fluctuations in a way that it produces an inward deuterium flux (figure \ref{fig:gk_lin_t0.3_nLi}). This effect is different from the collisionality effect described in \cite{fable}: Fable et al. showed that collisionality provides an outward contribution to the radial deuterium flux in a plasma with trace impurities, whereas in the present case larger lithium concentration leads to inward deuterium transport despite the increasing collisionality. 
However, if the collisionality is sufficiently high, the reduction of the impurity concentration towards the trace limit leads to outward main ion flux, in agreement with \cite{fable}.

\begin{figure}
 \begin{center}
  \includegraphics[scale=0.25]{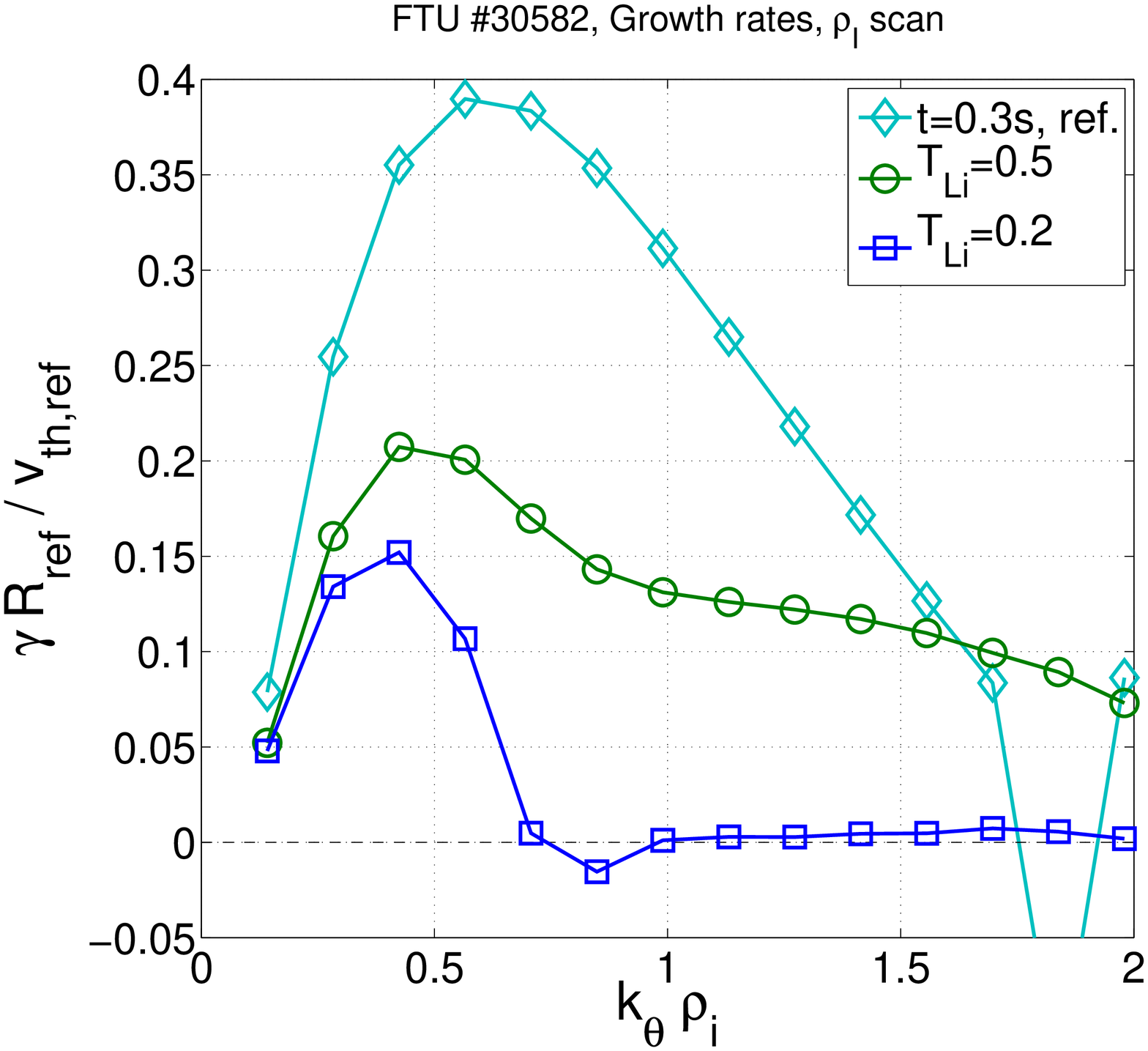}
  \includegraphics[scale=0.25]{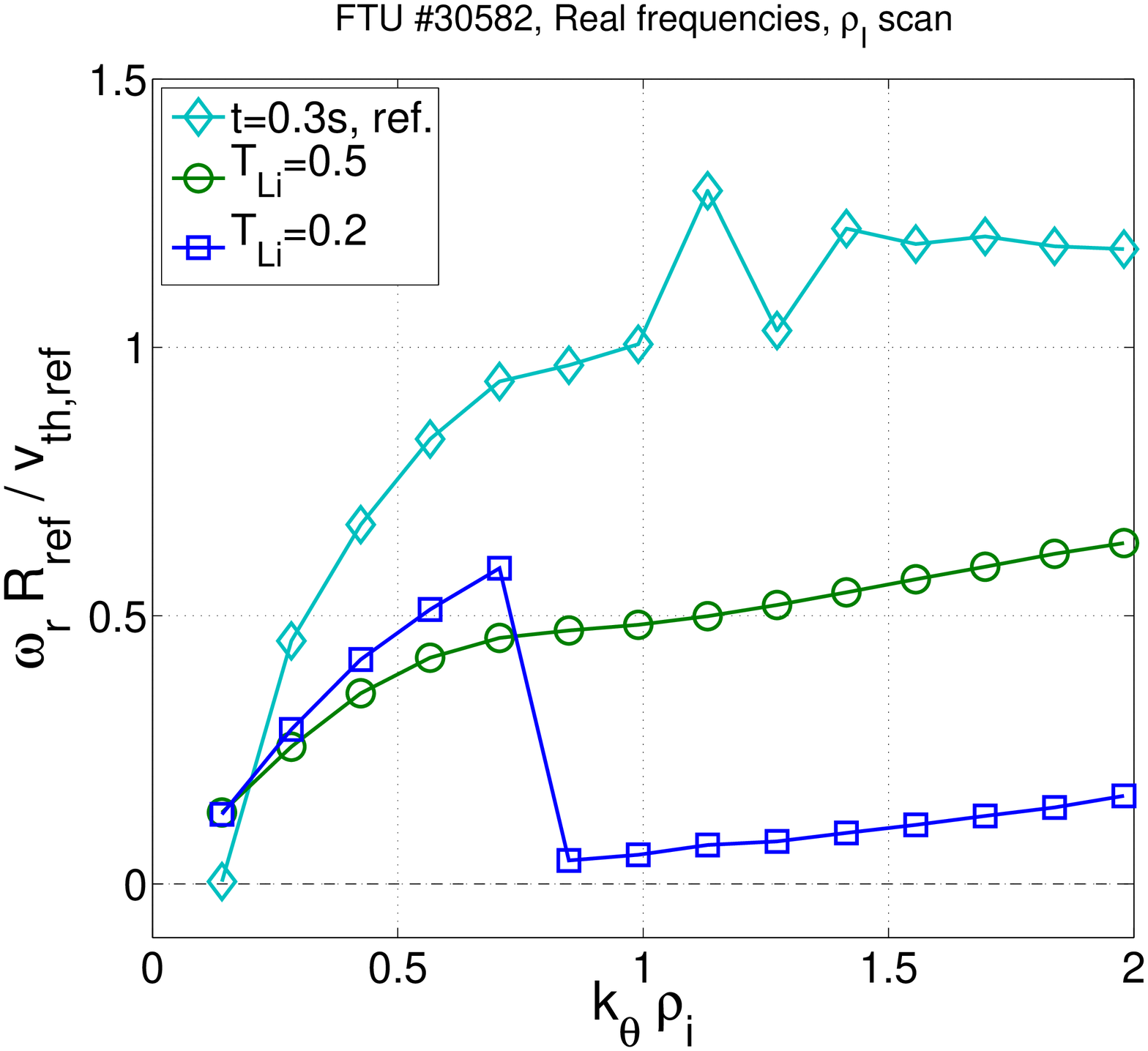} 
  \includegraphics[scale=0.25]{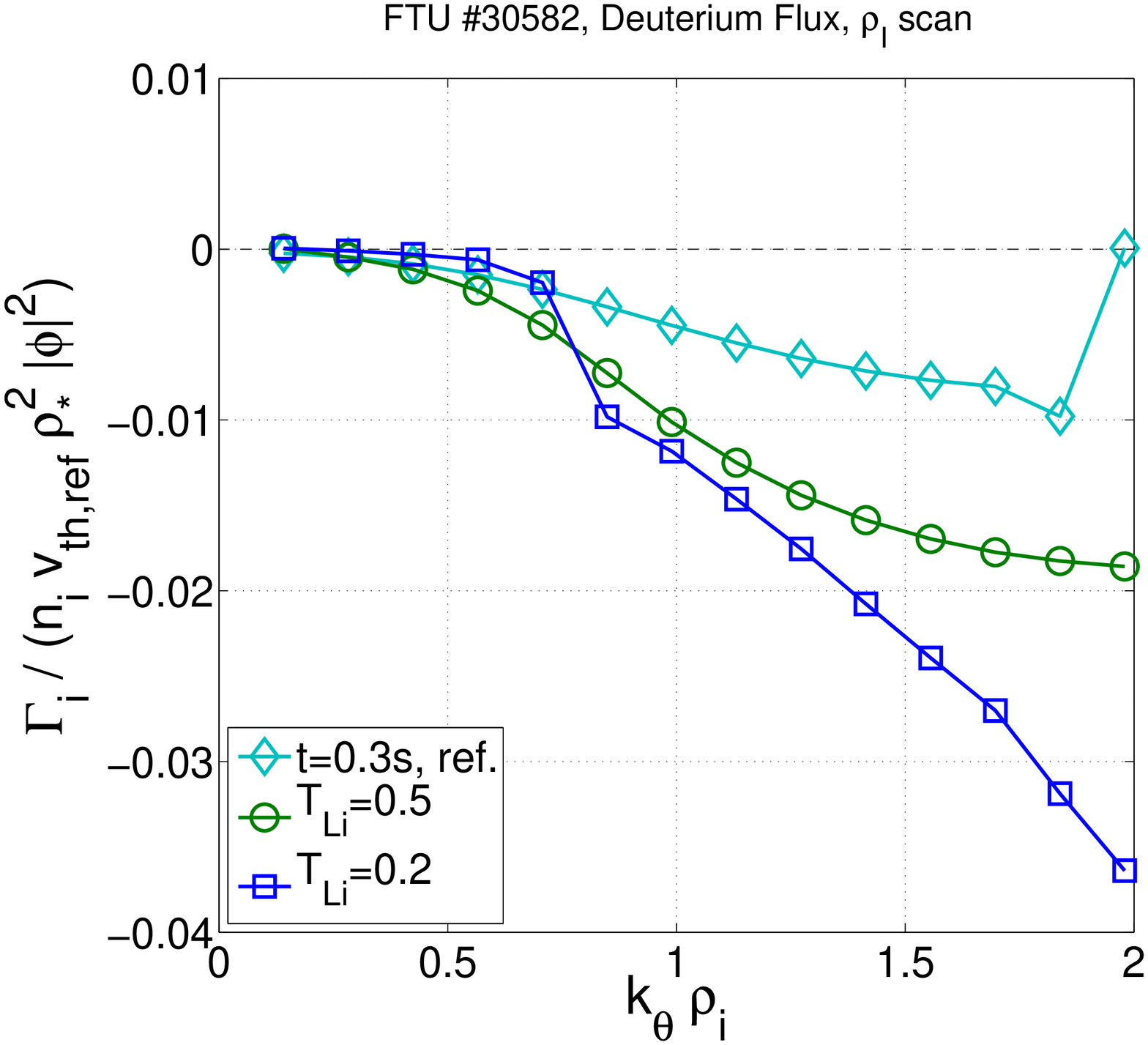}
  \includegraphics[scale=0.25]{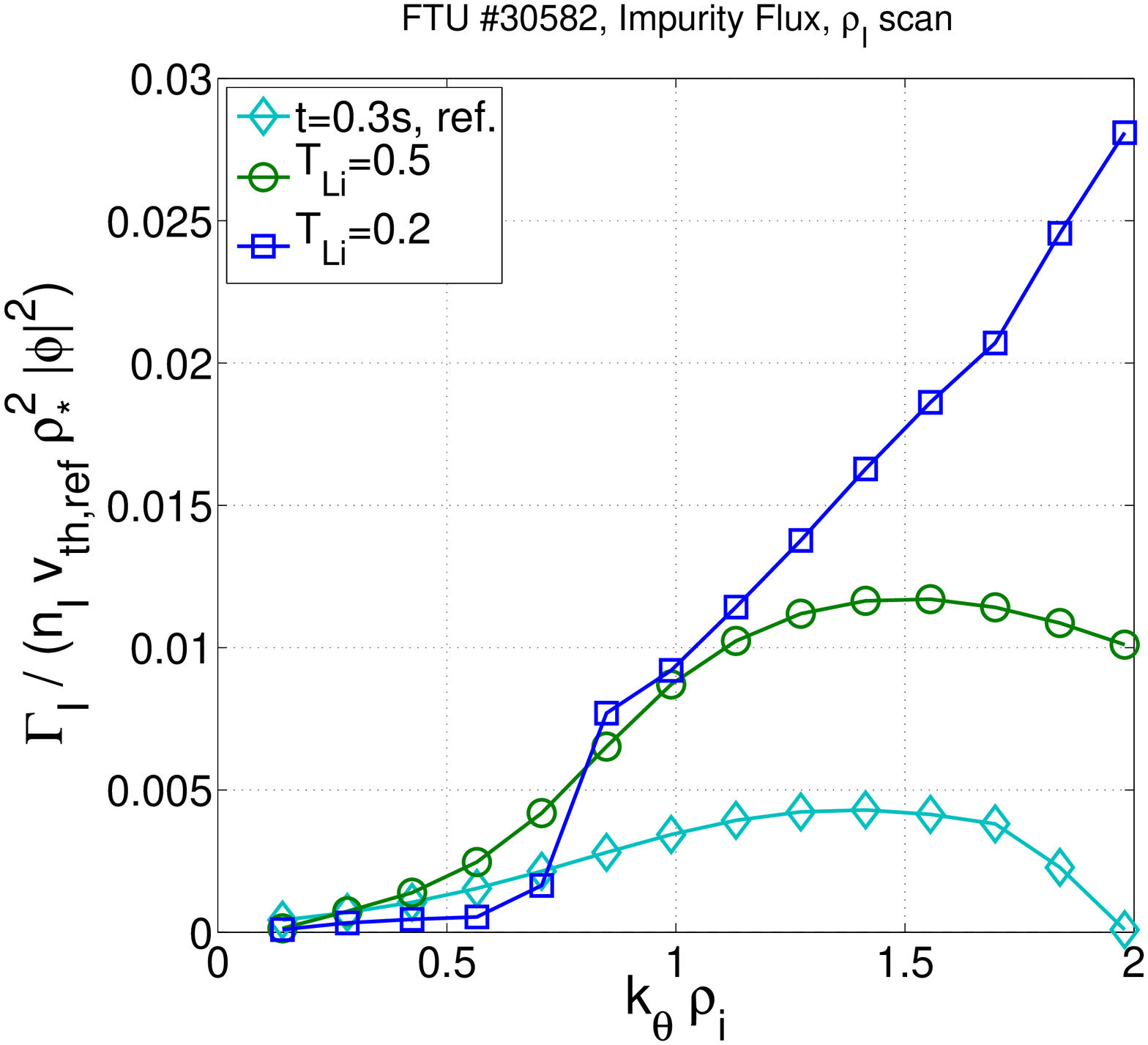}
 \end{center}
 \caption{Effect of changing the impurity Larmor-radius on the growth rate (top left), real frequency (top right), quasi-linear deuterium (bottom left) and lithium (bottom right) flux as a function of the bi-normal wavenumber at $t=0.3 \st{s}$.}
 \label{fig:gk_lin_t0.3_rhoI}
\end{figure}

The typical spatial scale of the impurity ITG modes can be separated from those of the deuterium by reducing the impurity temperature, and thus their Larmor-radius: $\rho_{\st{Li}} / \rho_{\st{D}} = Z_{\st{D}}/Z_{\st{Li}} \sqrt{m_{\st{Li}} T_{\st{Li}} / (m_{\st{D}} T_{\st{D}})}$. Figure \ref{fig:gk_lin_t0.3_rhoI} shows the effect of changing the $T_{\st{Li}}/T_{\st{D}}$ temperature ratio from 1 down to 0.5 (green circles) and 0.2 (blue squares). At 50\% reduced lithium temperature all ion modes are stabilized highlighting their mixed nature. The ion peak in the growth rate spectrum becomes wider as the modes at higher $k_{\theta}$ values are destabilized as a result of the smaller lithium Larmor-radius. A further cooling of the impurity species almost completely stabilizes the lithium modes due to their significantly increased collisionality. However, the peak of the lithium driven modes is now shifted even further away from the main D-ITG region. 
The phase difference between potential and ion density perturbations generated by the Li-ITG modes is larger than that by the D-ITG modes. However, the Li-ITG modes are located at higher wavenumber values, thus their contribution to the saturated flux is expected to be reduced. 

\subsection{Non-linear Analysis} \label{sec:gk_nonlin}
\begin{figure}
 \begin{center}
  \includegraphics[scale=0.25]{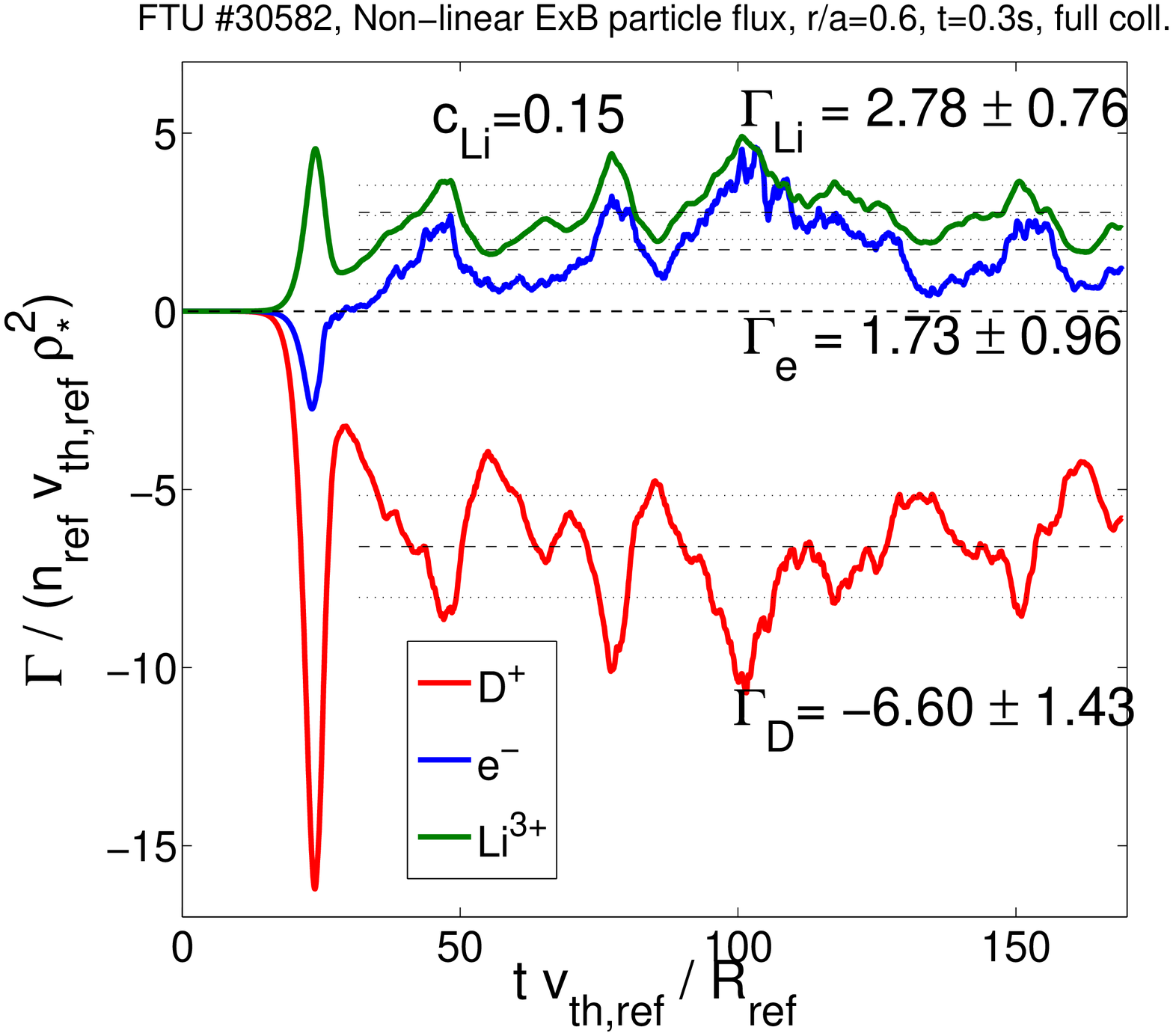}
  \includegraphics[scale=0.25]{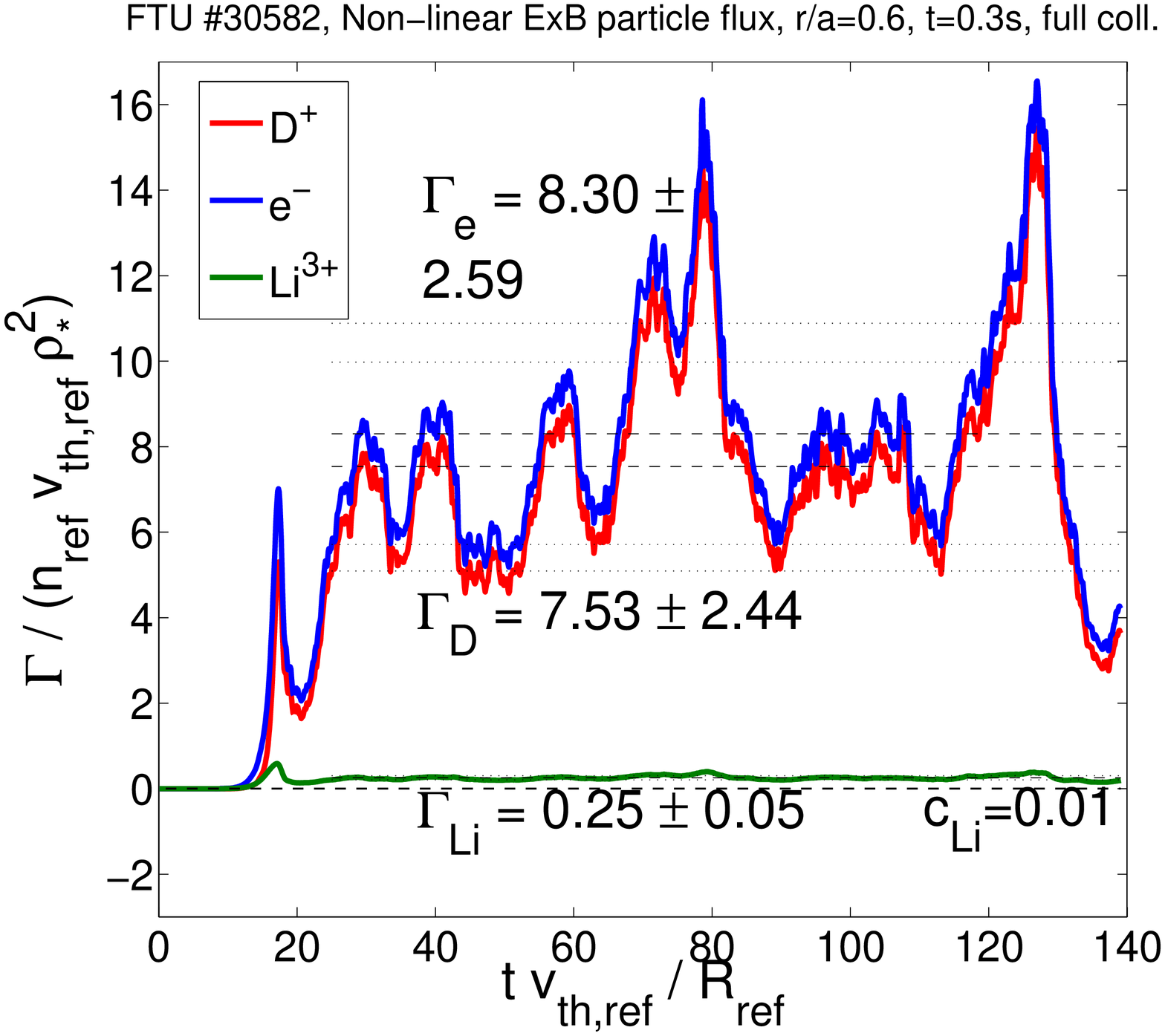} 
  \includegraphics[scale=0.25]{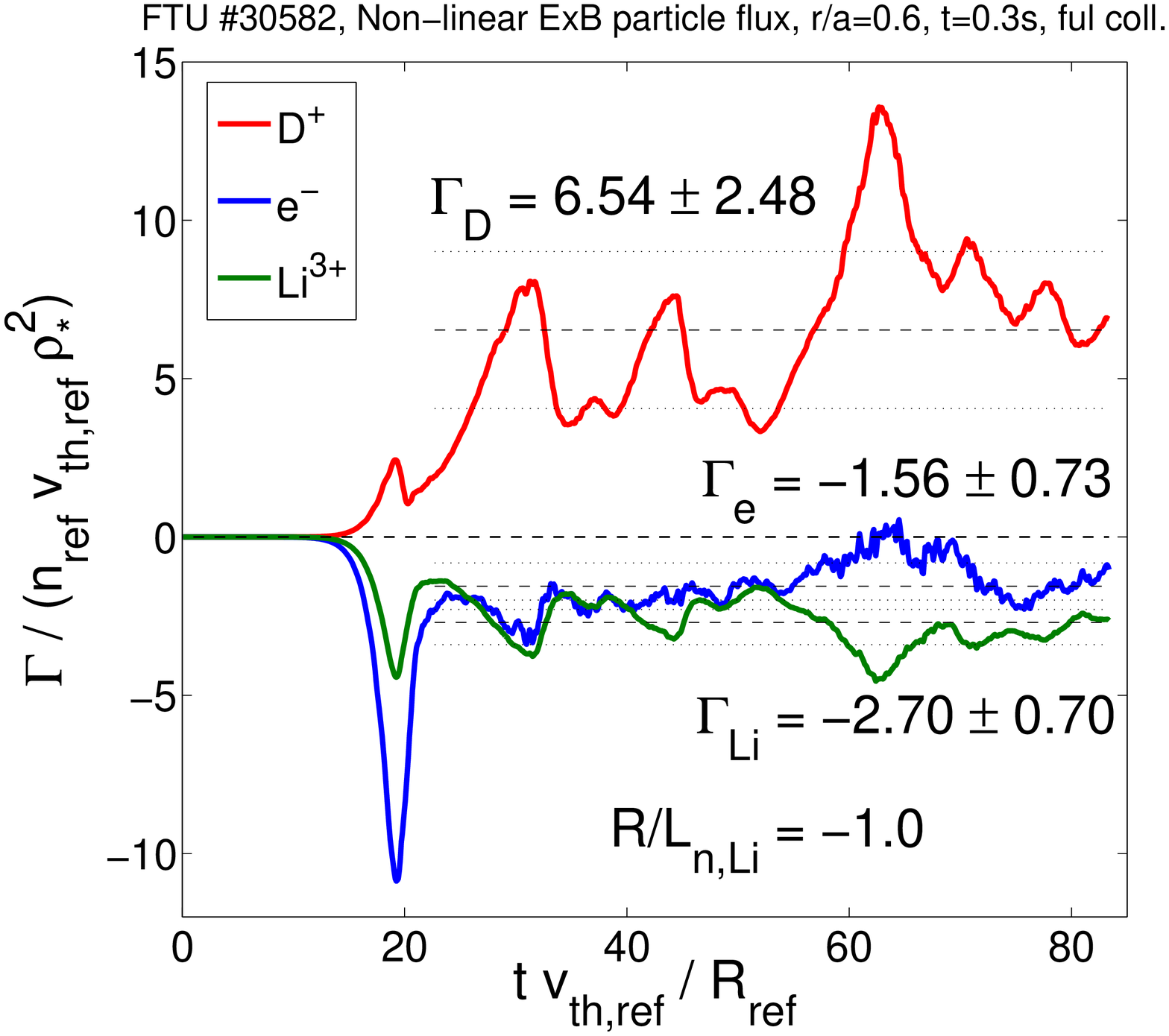}
  \includegraphics[scale=0.25]{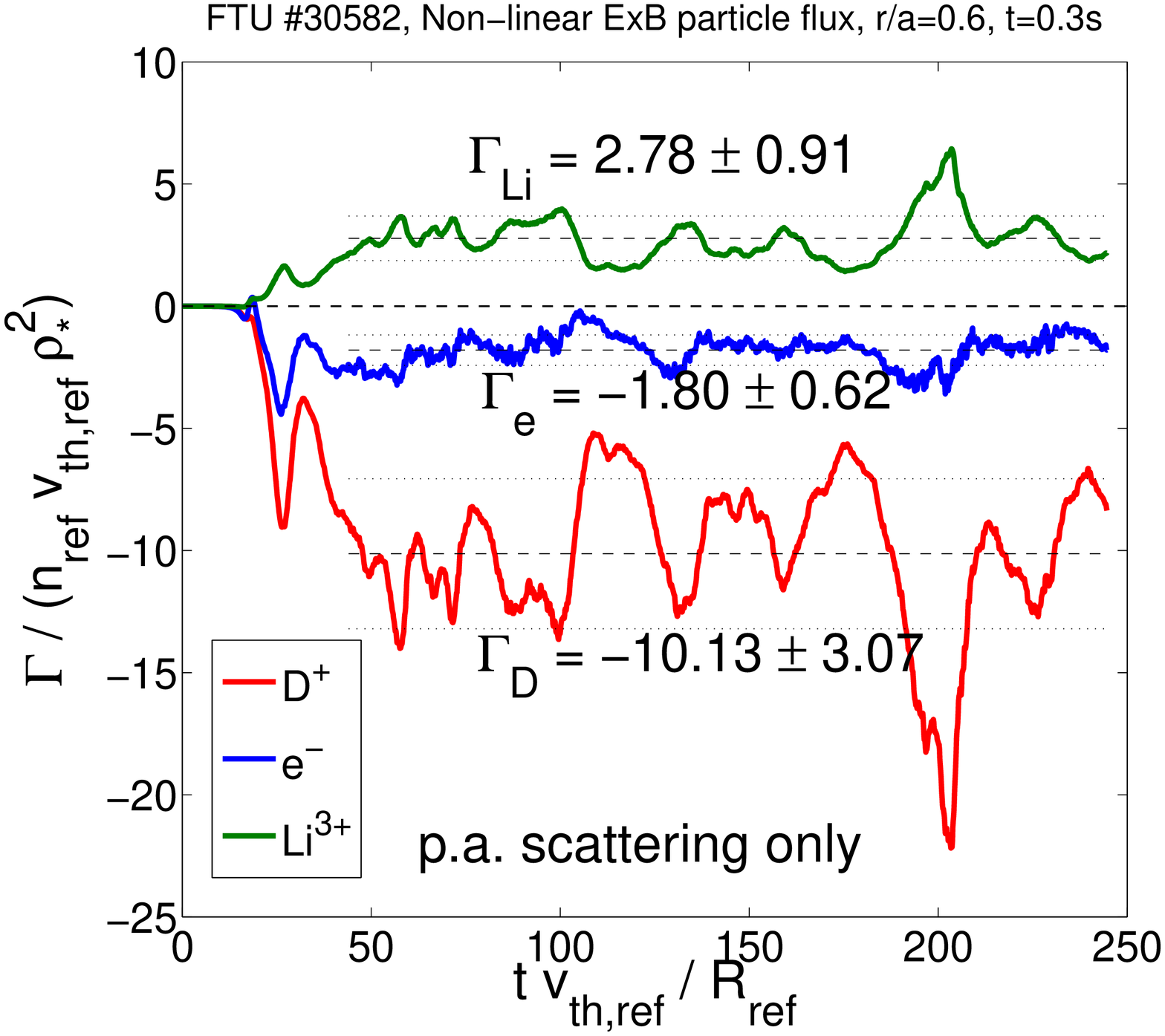}
 \end{center}
 \caption{Non-linear simulation of the radial particle flux driven by ITG modes as a function of time in FTU \#30582 at $t=0.3 \st{s}$ and $r/a=0.6$, using the full collision operator, with the experimental ($c_{\st{Li}}=0.15$, top left) and with reduced ($c_{\st{Li}}=0.01$, top right) lithium concentration. The experimental case with negative impurity density gradient ($R_{\st{ref}}/L_{\st{n,Li}}=-1.0$, bottom left) and with pitch-angle scattering only (bottom right) are also shown.}
 \label{fig:gk_nonlin_t0.3}
\end{figure}

A non-linear gyrokinetic study of the FTU \#30582 discharge has been carried out with GKW in order to support the linear gyrokinetic and fluid results. Figure \ref{fig:gk_nonlin_t0.3} shows four electro-static simulations of the radial particle flux of the species driven by the ITG modes as a function of time at $t=0.3 \st{s}$ and $r/a=0.6$. The top left panel is the reference, it corresponds to the experimental $c_{\st{Li}}=0.15$, while the top right panel to the reduced $c_{\st{Li}}=0.01$ impurity concentration. In both cases the full collision operator has been adopted. The simulation of the reference case has been performed with 21 bi-normal modes equally spaced up to $k_{\theta} \rho_\st{i}=1.6$ and 83 radial modes, yielding a box size of $\sim 75 \rho_{\st{i}}$ in both perpendicular directions. In the reduced lithium case the maximum wavenumber has been decreased to $k_{\theta} \rho_\st{i}=1.4$ in order to match the transition point from ITG to TEM in the corresponding linear spectrum. The resolution along the remaining dimensions is 16 points per period in parallel direction, 10 points in magnetic moment and 48 points in parallel velocity space. The fluxes are expressed in units of $n_{\st{ref}} v_{\st{th,ref}} \rho_*^2 \approx 7.9 \cdot 10^{18} \st{s}^{-1} \st{m}^{-2}$. 

The reference case (top left) shows that the deuterium and lithium flux maintain their directions as predicted by the linear analysis. However, the electron flux changes from inward to outward when progressing into the saturated phase. 
This suggests that the linear electron flux is determined by the fast growing modes above $k_{\theta} \rho_{\st{i}} \approx 0.5$ (see figure \ref{fig:gk_lin_t0.3_nLi}, top right) but they are overtaken by the low-k modes in the saturated phase. Therefore, quasi-linear methods for estimating electron transport have to be employed with caution in this particular case. 
The same non-linear simulation using pitch-angle scattering only (bottom right) gives approximately the same value for the lithium flux: $2.78 \pm 0.91$, and values reduced by approximately the same margin for both deuterium and electron fluxes: $-10.13 \pm 3.07$ and $-1.80 \pm 0.62$, respectively. This reduction is again in agreement with the expected outward contribution of the collisions \cite{fable}. 

If the impurity density gradient is negative on the analysed flux surface (figure \ref{fig:gk_nonlin_t0.3}, bottom left), the ion species flow in the opposite radial direction compared to the reference case, as anticipated from the linear results (figure \ref{fig:gk_lin_t0.3_RLnLi_nLi_scan}). The electron flux becomes negative due to the lower electron density gradient ($R_{\st{ref}}/L_{\st{n,e}}=1.1$). 

When the lithium concentration is reduced to $c_{\st{Li}}=0.01$ (figure \ref{fig:gk_nonlin_t0.3}, top right) both the deuterium and electron fluxes are outward and the lithium flux is negligible. The value of the outward electron flux in this case ($8.30 \pm 2.59$) is almost a factor of five larger than in the experimental case ($1.73 \pm 0.96$). The deuterium flux is thus reversed and the electron flux is significantly reduced by the high lithium concentration, as suggested by figure \ref{fig:gk_lin_t0.3_nLi} in the linear analysis. The heat fluxes of the species (not shown) are also reduced in this case by approximately a factor of two. 

\begin{figure}
 \begin{center}
  \includegraphics[scale=0.4]{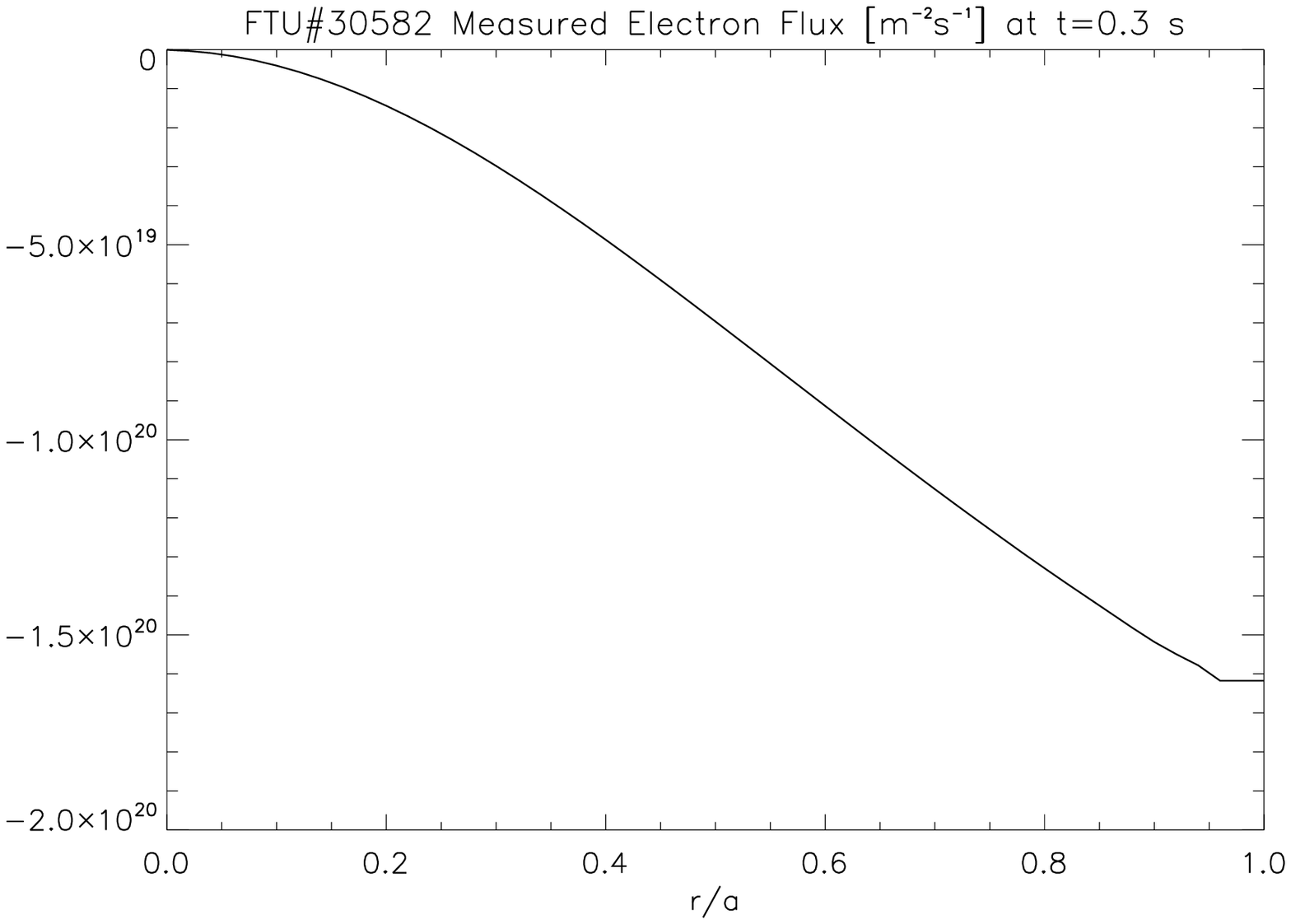}
  \includegraphics[scale=0.4]{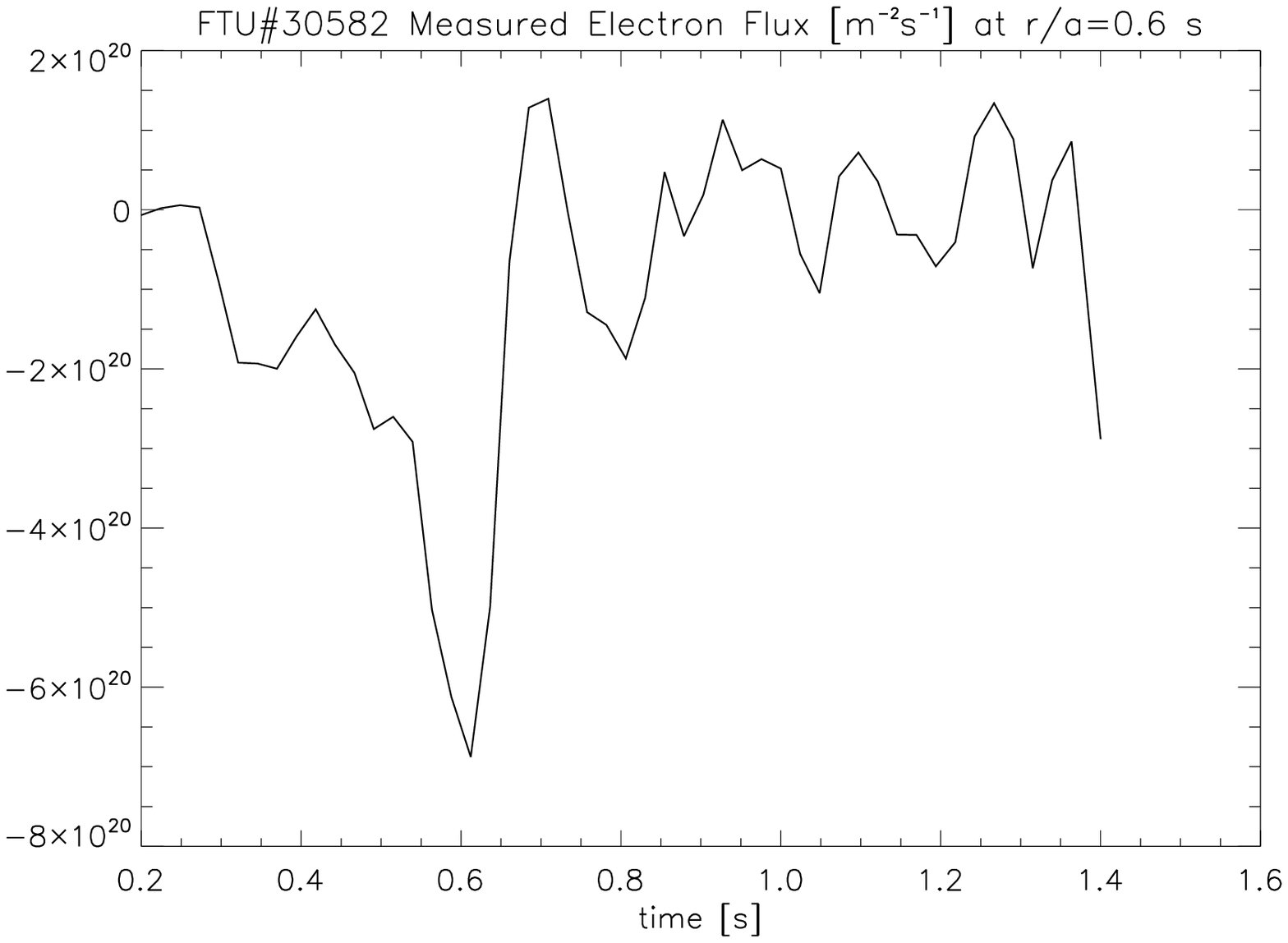} 
 \end{center}
 \caption{Estimate of the radial electron flux based on time evolution of the measured electron density profile. Left: flux profile at t=0.3s. Right: time traces of the flux at $r/a=0.6$. }
 \label{fig:el_transp_est}
\end{figure}

The electron flux can be estimated from the time evolution of the density profile by solving the transport equation $\partial_t n(\psi) = - \nabla_{\psi} \cdot \Gamma(\psi) + S(\psi)$, where $\psi$ is the radial coordinate. This calculation has been performed by the JETTO \cite{jetto} transport code assuming the source term $S$ to be zero, and the results have been plotted in figure \ref{fig:el_transp_est}. The left panel shows the experimental electron flux profile at $t=0.3$s, and the right panel the time traces of the flux at $r/a=0.6$. Since the estimated electron flux is much larger than the simulated one in the reference case (figure \ref{fig:gk_nonlin_t0.3}, top left), especially towards the edge, it suggests that there is significant electron source in the plasma at this time. However, the simulated electron flux in the reduced lithium concentration case is of the same magnitude as the estimated one, and thus it would effectively counteract the source term. The lithium induced electron flux reduction therefore seems to be an important factor in achieving a highly peaked electron density profile. The time traces of the estimated electron flux (figure \ref{fig:el_transp_est}, right) show that the electron density increases until $t=0.6$s, after which an oscillation around zero occurs in the stady-state phase of the discharge. The ion density profiles are not measured directly, the same analysis is not immediately available. However, since the average $Z_{\st{eff}}$ of the plasma decreases to near unity during the experiment, the deuterium profile must reach approximately the same degree of peaking as the electrons. 

Since FTU \#30582 is an Ohmically heated discharge with relatively low temperature and high magnetic field, in addition to the turbulent flux the Ware-pinch is expected to contribute to the total particle flux with a significant margin. The neoclassical particle transport has been estimated with the Neoart code \cite{neoart}. The turbulent electron flux predicted by GKW in physical units is $\Gamma_{\st{e}}^{\st{turb}} \approx 1.34 \cdot 10^{19} \st{s}^{-1} \st{m}^{-2}$. The overall neoclassical electron flux is calculated as $\Gamma_{\st{e}}^{\st{neo}} \approx -0.87 \cdot 10^{19} \st{s}^{-1} \st{m}^{-2}$ of which the flux due to the Ware-pinch is $\Gamma_{\st{e}}^{\st{Ware}} \approx -0.99 \cdot 10^{19} \st{s}^{-1} \st{m}^{-2}$. 
The turbulent and neoclassical values are thus comparable and the neoclassical contribution further reduces the electron transport. 
The total deuterium flux as estimated with the sum of the turbulent and neoclassical parts is $\Gamma_{\st{D}} \approx (- 5.21 - 0.43) \cdot 10^{19} \st{s}^{-1} \st{m}^{-2}$. 

The different direction of the simulated electron flux with the full collision operator and pitch-angle scattering only (figure \ref{fig:gk_nonlin_t0.3} top left and bottom right) is due to the fact that when the electron transport is close to zero. In this case the increased effect of collisions when using the full collision operator provides an additional outward component to the flux, and is sufficient to change the direction of the turbulent electron transport.

\section{Fluid Analysis} \label{sec:fluid}
The gyrokinetic results show that the high lithium content of the FTU-LLL plasmas contributes to the observed density peaking, and the inward deuterium flux is reversed when the impurities are removed from the system. They also indicate the importance of the centrally peaked impurity density profile and the similar lithium and deuterium Larmor-radii. In this section we introduce a fluid model that, although lacks the small scale physics of a gyrokinetic calculation, captures the main aspects of the ion particle transport and allows us to rapidly perform parameter scans. The fluid model describes all eigenmodes present in the system and it provides a simple way of separating the particle flux into diffusive, thermo-diffusive and pinch terms to assess their roles individually. While this is possible with gyrokinetic eigenvalue solvers, such as QuaLiKiz \cite{bourdelle_qualikiz} or GENE \cite{gene}, we decided to use a fluid model for its intuitive results. Our aim is to investigate the main drive of the lithium and deuterium transport when both of the ion species are centrally peaked. Although it has been shown that the electron transport can not be accurately modelled by quasi-linear estimates in this case, for the sake of completeness the electron flux values are also indicated. However, the focus of the fluid analysis is on the ion transport.

\subsection{The Model Equations}

The model is based on the collisionless Braginskii equations \cite{braginskii}, although collisions are introduced for the trapped electron species following the work by Nilsson and Weiland \cite{nilsson}. 
Collisions are required to stabilize TE modes and obtain the ITG driven transport.
Ion collisions have been neglected for simplicity. The model describes non-adiabatic main and impurity ions and trapped electrons. Passing electrons are considered adiabatic. The impurities are treated as non-trace species, i.e. their density response is taken into account in the quasi-neutrality equation. Magnetic perturbations are neglected (electrostatic). In the perpendicular momentum equation the drift approximation is applied, that is the fluctuating ExB and diamagnetic velocities are the leading order terms both estimated as $\mathcal{O}(v_{\st{th,s}} \rho_*)$. Gyro-viscous cancellation between the pressure tensor drift and the convective diamagnetic part of the polarization velocity is taken into account according to the result of Chang and Callen \cite{chang}, applicable in case of a non-uniform temperature profile. The FLR effects enter the equations through the polarization velocity. The system is closed by assuming that the heat flux is provided by the diamagnetic velocity (diamagnetic closure, \cite{weiland}). The model is radially local and only bi-normal and parallel fluctuations of the quantities are considered. 

After linearizing and Fourier-transforming, the continuity, parallel momentum and energy balance equations of the two ion species (denoted by subscript s) are
\begin{eqnarray}
 && \frac{\hat{n}_{\st{s,1}}}{n_{\st{s}0}} \left(2 \omega_{\st{D,s}} - \omega - \omega \frac{1}{2} k_{\st{y}}^2 \rho_{\st{th,s}}^2 \right) + \frac{\hat{T}_{\st{s,1}}}{T_{\st{s}0}} \left(2 \omega_{\st{D,s}} - \omega \frac{1}{2} k_{\st{y}}^2 \rho_{\st{th,s}}^2 \right) + \nonumber \\
 && \frac{Z_{\st{s}}e}{T_{\st{s}0}} \hat{\phi}_1 \left( 2 \omega_{\st{D,s}} - \omega_{*\st{s}}^{\st{n}} - \omega \frac{1}{2} k_{\st{y}}^2 \rho_{\st{th,s}}^2 \right) + k_{\parallel} \hat{v}_{\parallel \st{s}}^1 \left(1 + \frac{1}{2} k_{\st{s}}^2 \rho_{\st{th,s}}^2 \right) = 0
 \label{eq:continuity_lin}
\end{eqnarray}
\begin{eqnarray}
 - \frac{m_{\st{s}}}{T_{\st{s}}} \omega \hat{v}_{\parallel \st{s}}^1 + k_{\parallel} \left( \frac{\hat{n}_{\st{s}1}}{n_{\st{s}0}} + \frac{\hat{T}_{\st{s}1}}{T_{\st{s}0}} + \frac{Z_{\st{s}} e}{T_{\st{s}0}} \hat{\phi}_1 \right) = 0
 \label{eq:par_momentum_lin}
\end{eqnarray}
\begin{eqnarray}
 \left( \frac{10}{3} \omega_{\st{D,s}} - \omega \right) \frac{\hat{T}_{\st{s}1}}{T_{\st{s}0}} + \frac{2}{3} \omega \frac{\hat{n}_{\st{s}1}}{n_{\st{s}0}} - \frac{Z_{\st{s}}e}{T_{\st{s}}} \hat{\phi}_1 \omega_{*\st{s}}^{\st{n}} \left( \eta_{\st{s}} - \frac{2}{3} \right) = 0
 \label{eq:energy_lin}
\end{eqnarray}
where $\hat{n}_{\st{s,1}}$, $\hat{T}_{\st{s}1}$, $\hat{v}_{\parallel \st{s}}^1$ and $\hat{\phi}_1$ are the Fourier components of the density, temperature, parallel velocity and electrostatic potential fluctuations, respectively, the corresponding quantities with a 0 subscript denote the equilibrium values. $k_{\parallel}$ and $k_{\st{y}}$ are the parallel and bi-normal wavenumbers, $\rho_{\st{th,s}}=(m_{\st{s}} v_{\st{th,s}})/(Z_{\st{s}} e B)$ is the species' thermal Larmor-radius. $\omega$ is the mode frequency, $2 \omega_{\st{D,s}} = k_{\st{y}} \vec{y} \cdot \vec{v}_{\st{D,s}}$ is the magnetic drift frequency and $\omega_{*\st{s}}^{\st{n}} = - k_{\st{y}} T_{\st{s}} / (Z_{\st{s}} e B L_{\st{n,s}})$ is the diamagnetic frequency, $L_{\st{n,s}} = -\frac{n_{\st{s}}}{\nabla n_{\st{s}}}$ is the density gradient length scale and $\eta_{\st{s}} = L_{\st{n,s}} / L_{\st{T,s}}$ is the ratio of the density and temperature gradient length scales.

The bounce-averaged parallel motion of the trapped electrons and the electron Finite Larmor Radius (FLR) terms are neglected. The trapped electron continuity equation is
\begin{eqnarray}
 \frac{\hat{n}_{\st{e,1}}}{n_{\st{et}0}} \left(2 \omega_{\st{D,e}} - \omega - \st{i} \nu_{\st{th}} \right) + \frac{\hat{T}_{\st{e,1}}}{T_{\st{e}0}} 2 \omega_{\st{D,e}} + \frac{Z_{\st{e}}e}{T_{\st{e}0}} \hat{\phi}_1 \left( 2 \omega_{\st{D,e}} - \omega_{*\st{e}}^{\st{n}} - \st{i} \nu_{\st{th}} \Gamma \right) = 0
 \label{eq:continuity_dtem}
\end{eqnarray}
and their energy balance equation is
\begin{eqnarray}
 && \left( \frac{10}{3} \omega_{\st{D,e}} - \omega \right) \frac{\hat{T}_{\st{e}1}}{T_{\st{e}0}} + \frac{\hat{n}_{\st{e}1}}{n_{\st{et}0}} \left( \frac{2}{3} \omega  - \beta \st{i} \nu_{\st{th}} \right) - \nonumber \\ 
 && \frac{Z_{\st{e}}e}{T_{\st{e}}} \hat{\phi}_1 \left[ \omega_{*\st{e}}^{\st{n}} \left( \eta_{\st{e}} - \frac{2}{3} \right) + \beta \st{i} \nu_{\st{th}} \right] = 0
 \label{eq:energy_dtem}
\end{eqnarray}
where $n_{\st{et}0}=n_{\st{e}} f_{\st{t}}$ and the trapped electron fraction is calculated as $f_{\st{t}}=\sqrt{2 r / (r + R)} \approx 0.55$ in this case. $\nu_{th} = \nu_{\st{e}} / \epsilon$, $\epsilon = r/R$ the inverse aspect ratio, $\Gamma = 1 + \frac{\alpha \eta_{\st{e}} \omega_{*\st{e}}^{\st{n}}}{\omega - \omega_{\st{D,e}} + \st{i} \nu_{\st{th}}}$. $\alpha \approx 1$ and $\beta \approx 1.5$ are factors determined in \cite{nilsson} in order to recover the strongly collisional TE response with the simplified collision operator. 

The effect of magnetic geometry is taken into account according to Hirose \cite{hirose}: The norms of the parallel and perpendicular differential operators are estimated by taking the average of an ad-hoc strongly ballooning eigenfunction in the ballooning space, resulting $k_{\parallel}=1/\sqrt{3 (q R)^2}$, $k_{\perp}=k_{\st{y}} \sqrt{1+(\pi^2/3-5/2)\hat{s}^2}$ replacing $k_{\st{y}}$ in the FLR terms, $\omega_{\st{D,s}}=-k_{\st{y}} T_{\st{s}0} / (Z_{\st{s}} e R) (2/3 + 5/9 \hat{s})$ and $\omega_{\st{D,e}}=-k_{\st{y}} T_{\st{e}0} / (Z_{\st{e}} e R) (1/4 + 2/3 \hat{s})$ in the ion and trapped electron magnetic drift frequencies, respectively \cite{nordman}. $R$ is the tokamak major radius, $q$ is the safety factor, $\hat{s}=r/q\ \st{d}q / \st{d}r$ is the magnetic shear. This model is similar to the one used by Moradi et al. \cite{moradi} for impurity transport studies. The main difference is in the treatment of the FLR terms: here we keep the FLR corrections separate so they appear in the density, temperature parallel velocity and potential terms, as well, while Moradi et al. introduce a flute mode solution providing an equation between the pressure and potential perturbations \cite{tardini}. The difference between the results of the two models is negligible. 

The above equations together with the quasi-neutrality condition $ \sum_{\sigma} Z_{\sigma} n_{\sigma 1} = 0$ ($\sigma$ indexes all four species) gives a closed set of equations. The dispersion relation obtained from it is a ninth degree polynomial in the complex mode frequency $\omega$ and it is evaluated numerically with Matlab.

The particle flux is expressed as $\Gamma_{\st{\sigma,r}} = \langle n_{\st{\sigma}1} \vec{r} \cdot \vec{v}_{\st{\sigma}1} \rangle$ where the angled brackets denote flux-surface averaging. The saturated value of the electrostatic potential required to evaluate this formula is estimated by the Weiland-model assuming an isotropic turbulent state: $|\phi| = \gamma B / (k_{\st{y}})$ \cite{weiland}. The particle flux can formally be split into terms explicitly proportional to the normalized density and temperature gradients ($R_{\st{ref}}/L_{\st{n,\sigma}}$ and $R_{\st{ref}}/L_{\st{T,\sigma}}$) of the species, and a residual term \cite{angioni_rev}. These contributions are typically called diffusive, thermo-diffusive and pinch terms, and the flux can be expressed as
\begin{eqnarray}
 \Gamma_{\st{\sigma,r}}^k = \frac{n_{\st{\sigma},0}}{R_{\st{ref}}} \left( D_{\st{n,\sigma}} \frac{R_{\st{ref}}}{L_{\st{n,\sigma}}} + D_{\st{T,\sigma}} \frac{R_{\st{ref}}}{L_{\st{T,\sigma}}} + R_{\st{ref}} V_{\st{p,\sigma}} \right)
\end{eqnarray}
where $D_{\st{n,\sigma}}$ is the diffusion coefficient, $D_{\st{T,\sigma}}$ the thermo-diffusion coefficient and $V_{\st{p,\sigma}}$ the pinch velocity. Note, however, that this separation is purely formal (unlike for the impurity transport in the trace impurity limit): due to the dependence of the mode frequency on the plasma parameters, the above coefficients depend on the gradients themselves, and in general the transport cannot be treated as a linear problem. Nonetheless, these terms provide useful additional information about the main driving mechanisms of particle transport. 
In the present description the term proportional to the magnetic shear, explicitly appearing in equation 2 in \cite{romanelli}, is incorporated in the residual term. However, all three components of the flux are separated into curvature and slab contributions. The curvature pinch term is related to the magnetic shear term in \cite{romanelli} but it is not clearly proportional to $\hat{s}$ and therefore a direct comparison is not possible.

\subsection{Fluid Results}
In this section we present the analysis of the two time instances with the fluid model and compare the results with those in section \ref{sec:gk_lin}. 

\begin{figure}
 \begin{center}
  \includegraphics[scale=0.25]{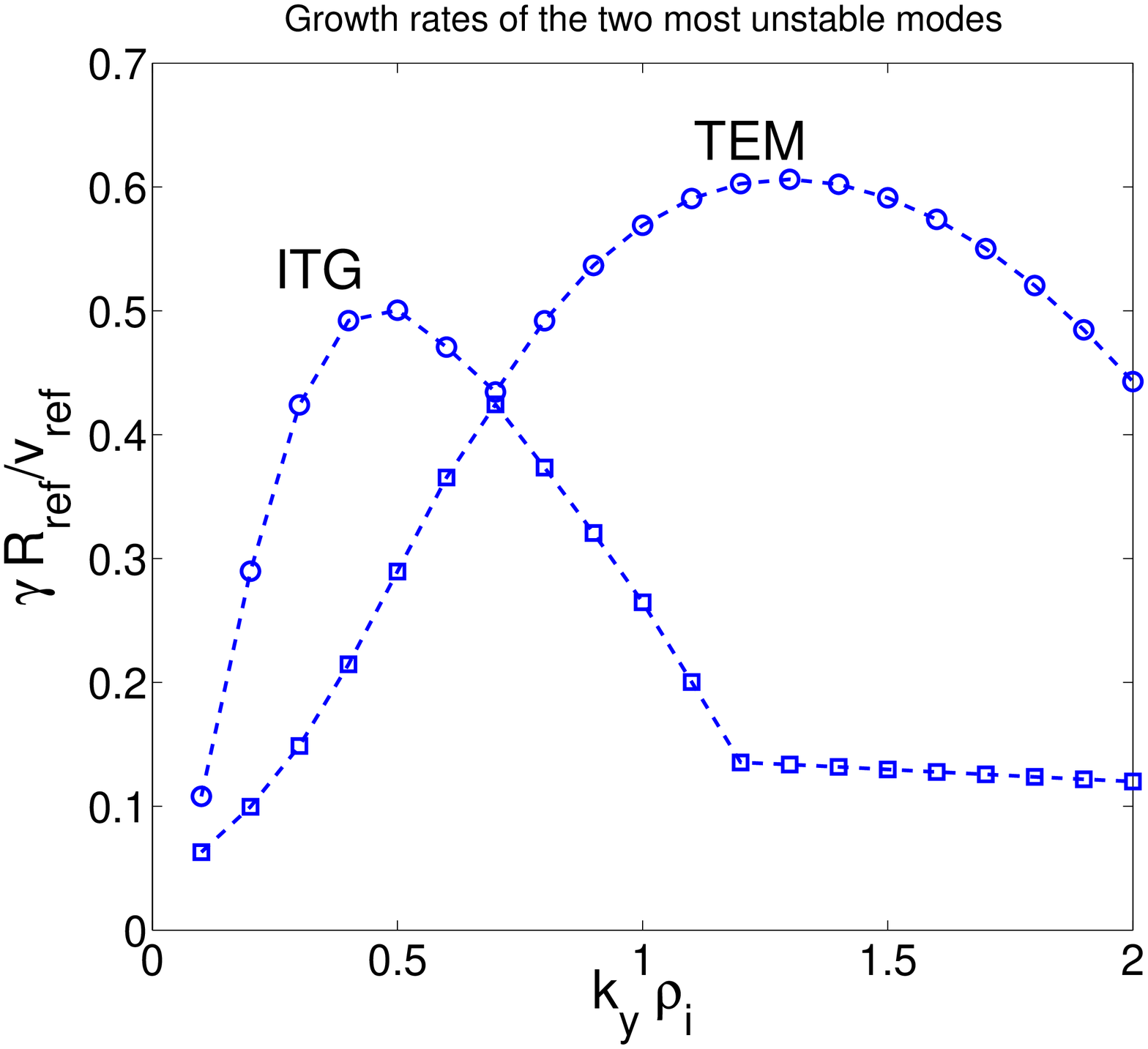}
  \includegraphics[scale=0.25]{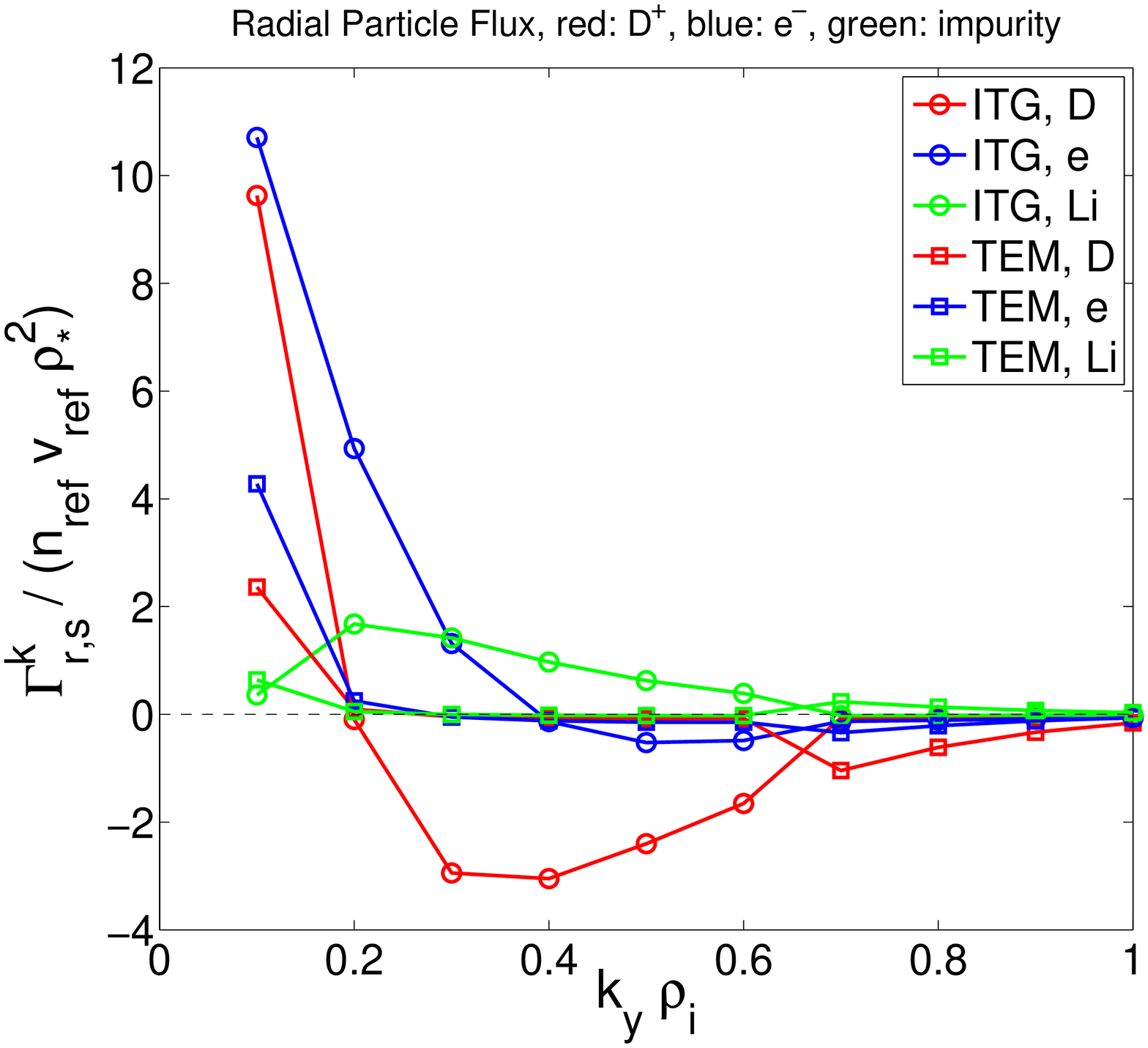} 
  \includegraphics[scale=0.25]{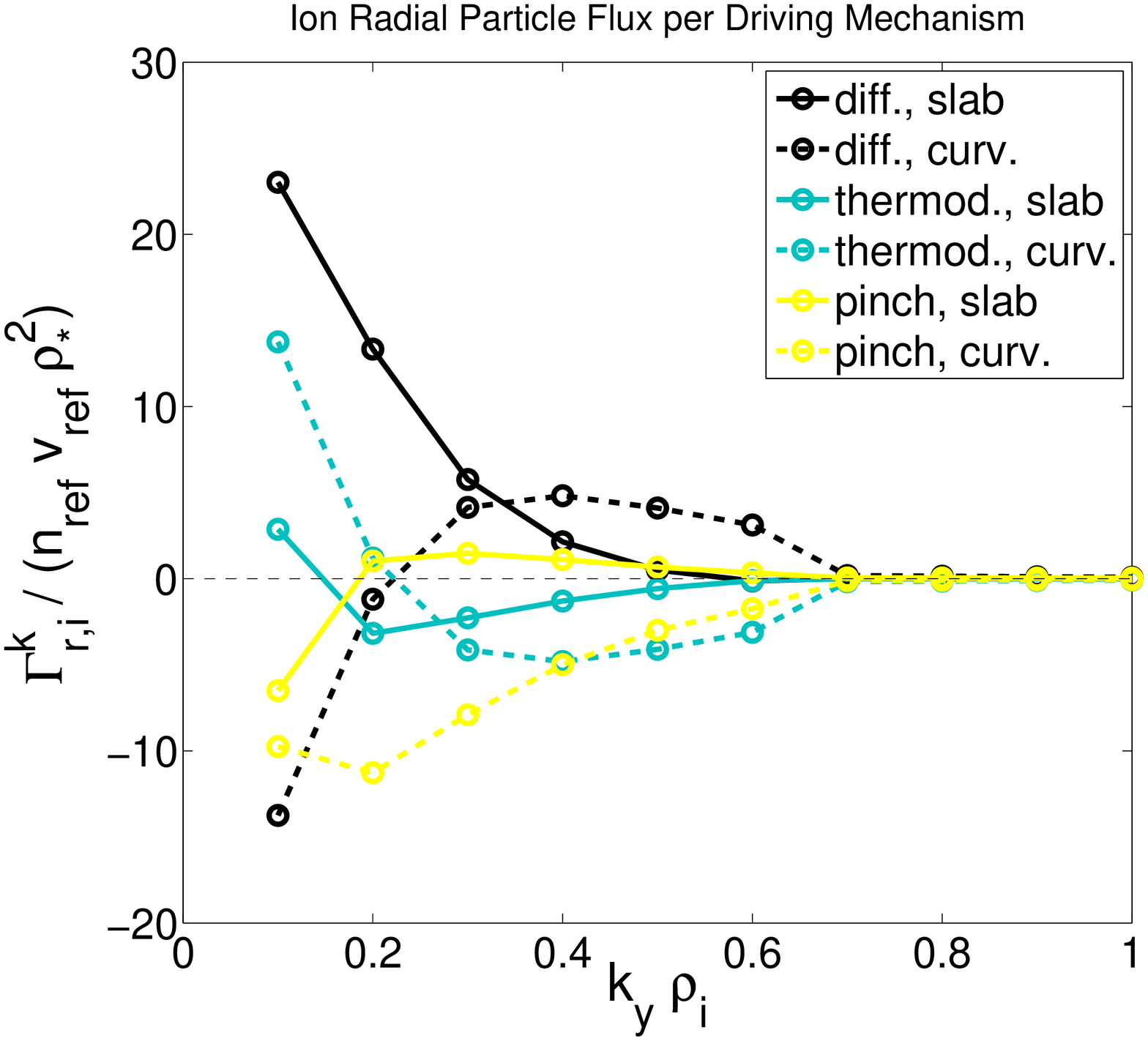}
  \includegraphics[scale=0.25]{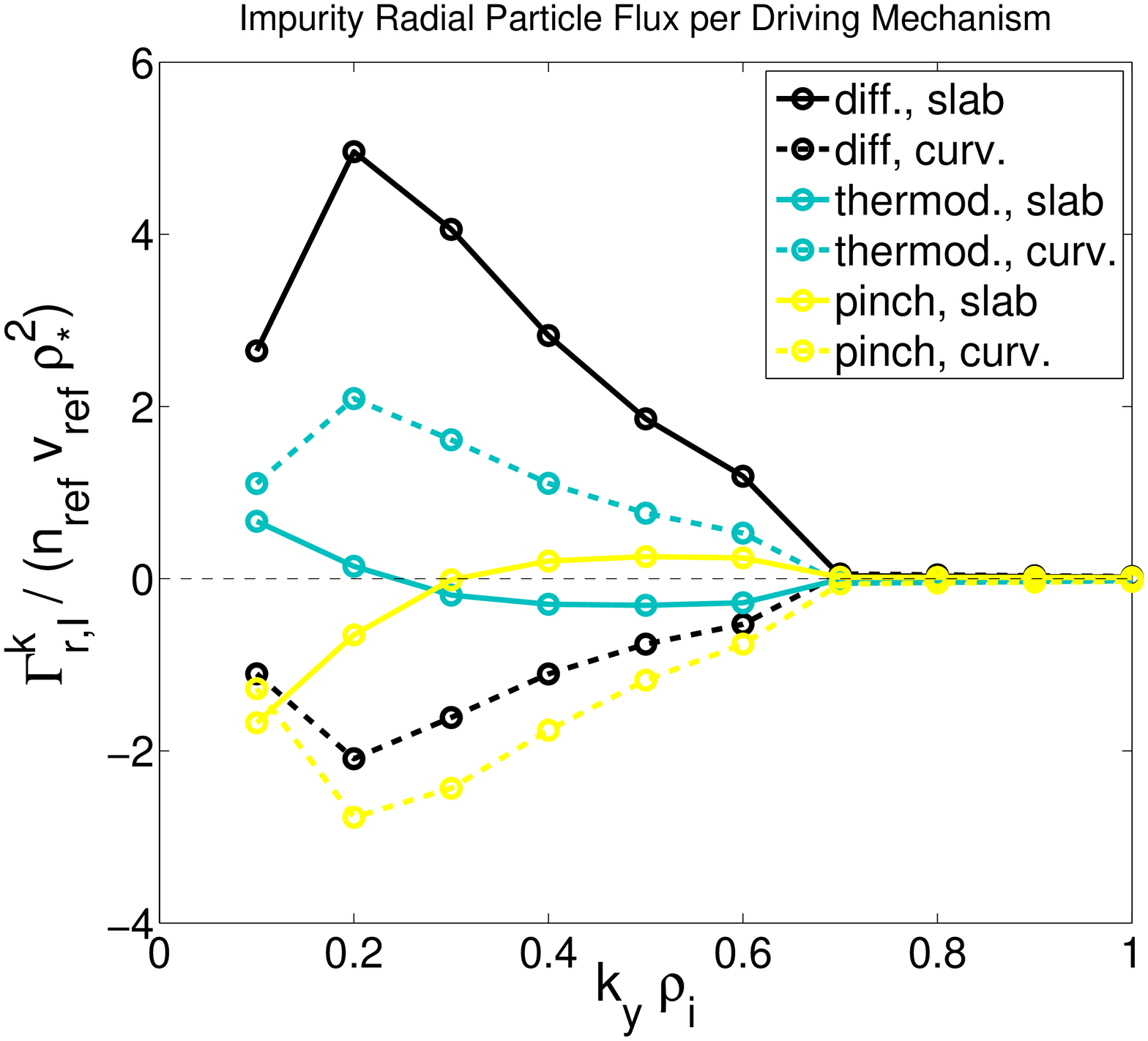}
 \end{center}
 \caption{Fluid analysis at $t=0.3 \st{s}$ with the reference $c_{\st{Li}}=0.15$ lithium concentration. Growth rates of the two most unstable modes (ITG and TEM, top left), total flux of the species driven by ITG and TE modes (top right), deuterium (bottom left) and lithium (bottom right) particle flux by slab (solid) and curvature (dashed) terms of the diffusive (black), thermo-diffusive (cyan) and pinch (yellow) contributions driven by ITG modes as a function of the bi-normal wavenumber.}
 \label{fig:fluid_qlin_t0.3}
\end{figure}

Figure \ref{fig:fluid_qlin_t0.3} shows the growth rate of the two most unstable modes (top left), total particle flux of the three species (top right), deuterium (bottom left) and lithium (bottom right) fluxes driven by the ITG modes separated to slab and curvature terms of the diffusive, thermo-diffusive and pinch contributions as a function of the bi-normal wavenumber, with the parameters of the $t=0.3 \st{s}$ case. 
The ITG-s are mixed deuterium-lithium modes with only one eigenmode associated to them. Except for the lowest $k_{\theta}$ mode, the deuterium flux driven by the ion modes is directed inward across the spectrum. 
Compared with the gyrokinetic simulations (figure \ref{fig:gk_lin_t0.3_nu}), these two observations suggest that the effect of collisionality in the fluid model is underestimated at high, and overestimated at low $k_{\st{y}}$ values due to the $1/k_{\st{y}}$ scaling of the collision operator. The reason for the inward deuterium transport in this case is that the strong outward slab diffusive term is compensated by the the inward curvature pinch in the $0.2<k_{\theta} \rho_{\st{i}}<0.7$ range, as seen on the bottom left panel of the figure. The apparent cancellation between the diffusive and thermo-diffusive curvature terms is a coincidence, taking place only at the present gradient values. The outward lithium transport is dominated by the slab diffusive contribution across the spectrum. 

Although a precise quantitative agreement between the fluid and gyrokinetic results is not expected, the ratio of the deuterium and lithium flux in the linear gyrokinetic, non-linear gyrokinetic and quasi-linear fluid calculations at $k_{\st{y}} \rho_{\st{i}} = 0.4$ are all between minus one half and minus one-third. The coefficient $C_{\st{T,\sigma}}=D_{\st{T,\sigma}}/D_{\st{n,\sigma}}$, as calculated from the fluid model, takes a value between approximately -0.28 and -0.43 for deuterium ions in the region $0.3 < k_{\theta} \rho_{\st{i}} < 0.5$, where the largest flux is observed. These values are similar to those found in \cite{romanelli} for a metallic limiter FTU scenario. 

If the collision frequency is sufficiently low, TEM-s become the dominant mode, but the ITG driven deuterium flux remains to be directed inward. This result can also be seen in an ITG dominated regime obtained by artificially reducing the trapped electron fraction.

The effect of reducing the lithium concentration of the plasma to $c_{\st{Li}}=0.01$ at the $t=0.3 \st{s}$ case is that the slab diffusive component stays dominant in a wider region of the $k_{\theta}$ space, resulting in a strong outward electron and deuterium flux up until $k_{\theta} \rho_{\st{i}} \approx 0.5$ and a weak inward flux above.

\begin{figure}
 \begin{center}
  \includegraphics[scale=0.25]{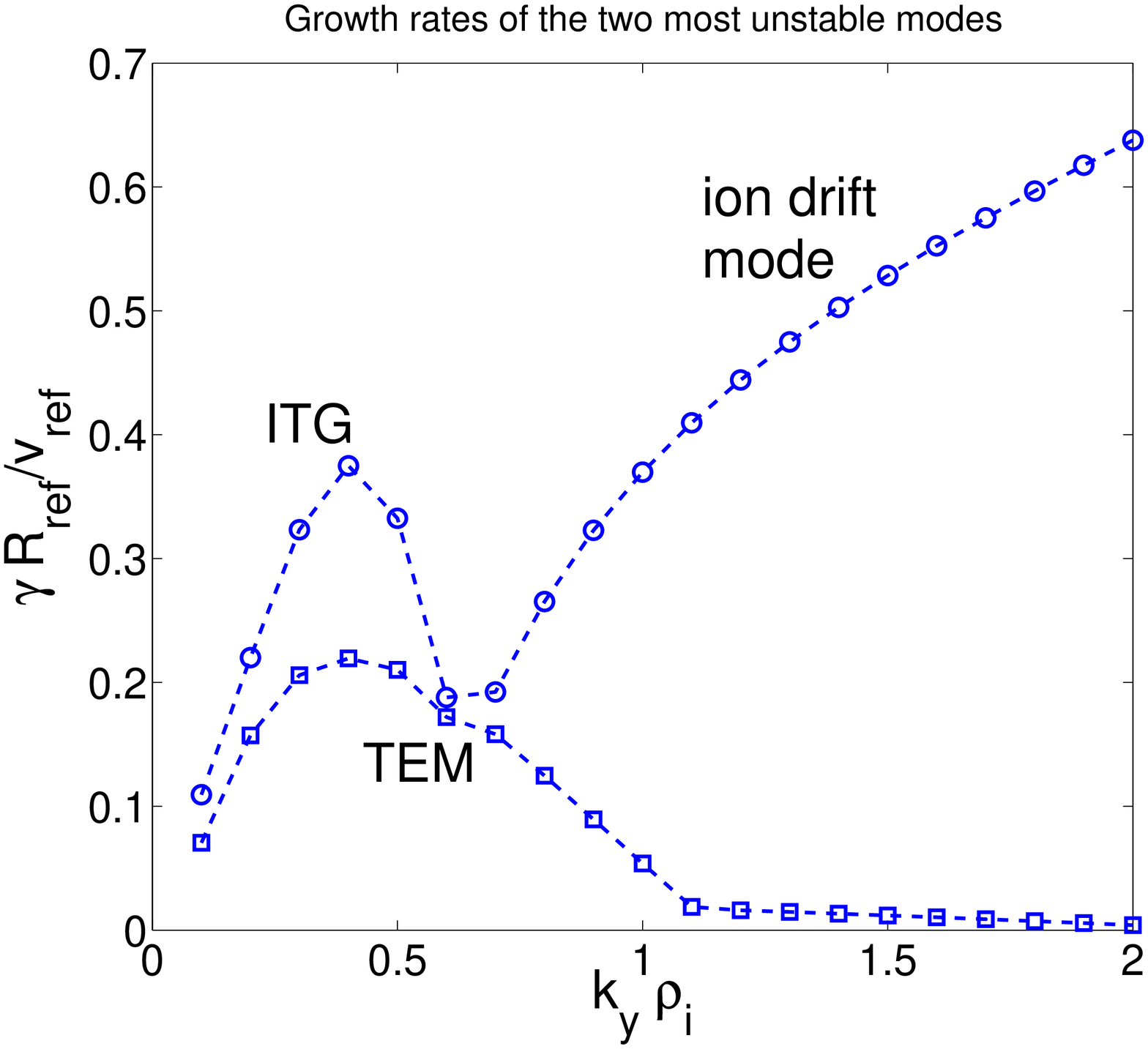}
  \includegraphics[scale=0.25]{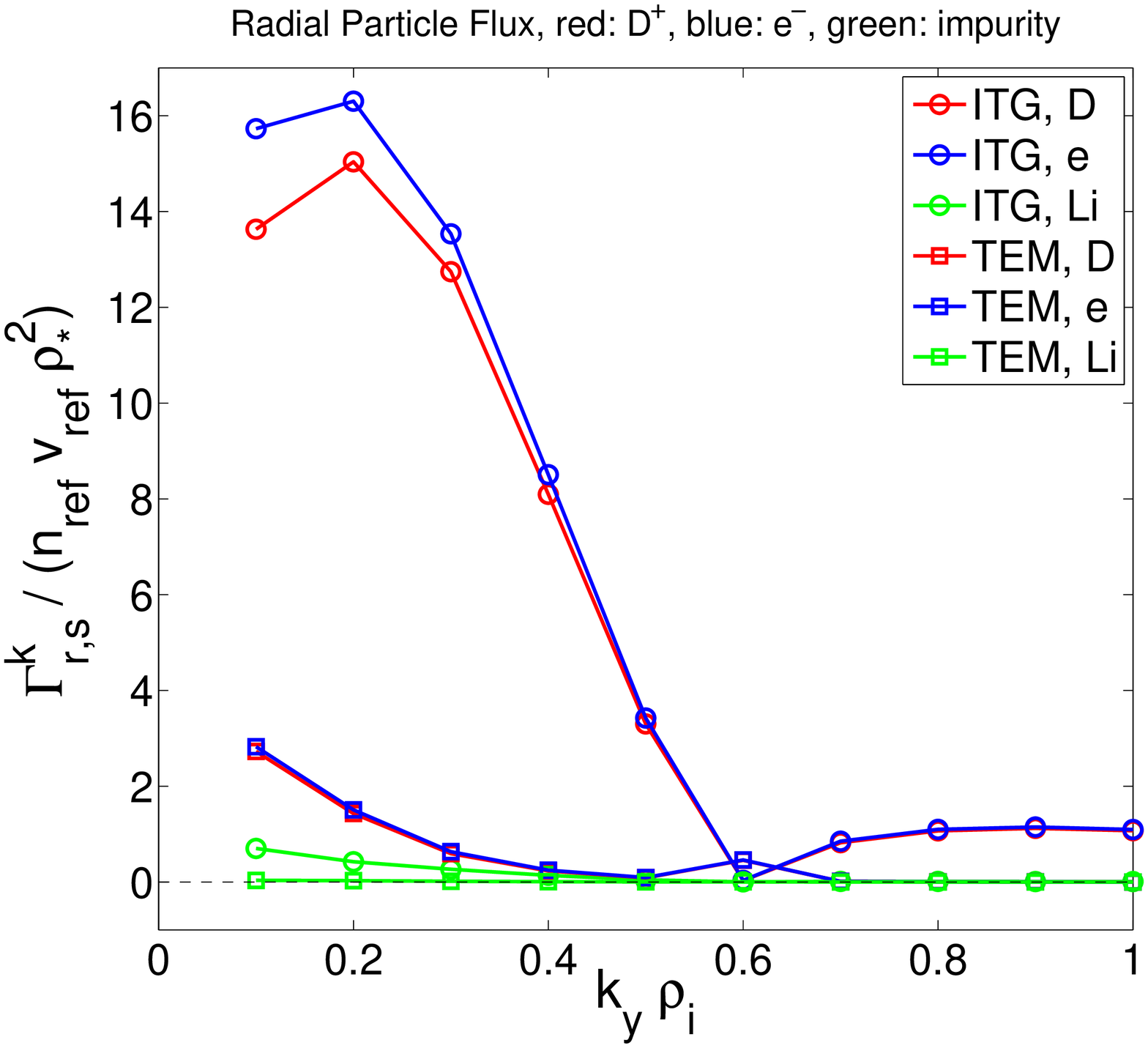} 
  \includegraphics[scale=0.25]{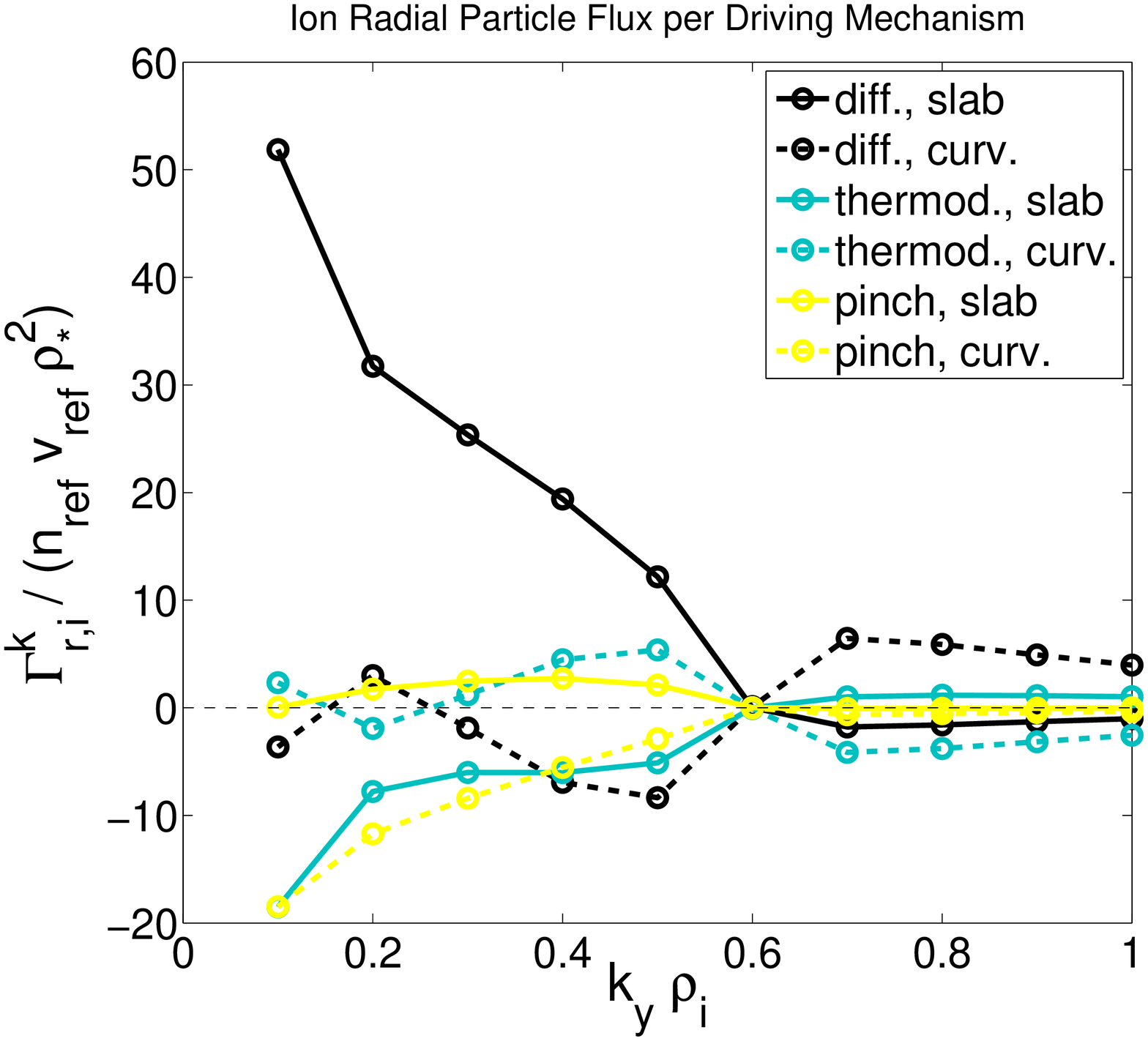}
  \includegraphics[scale=0.25]{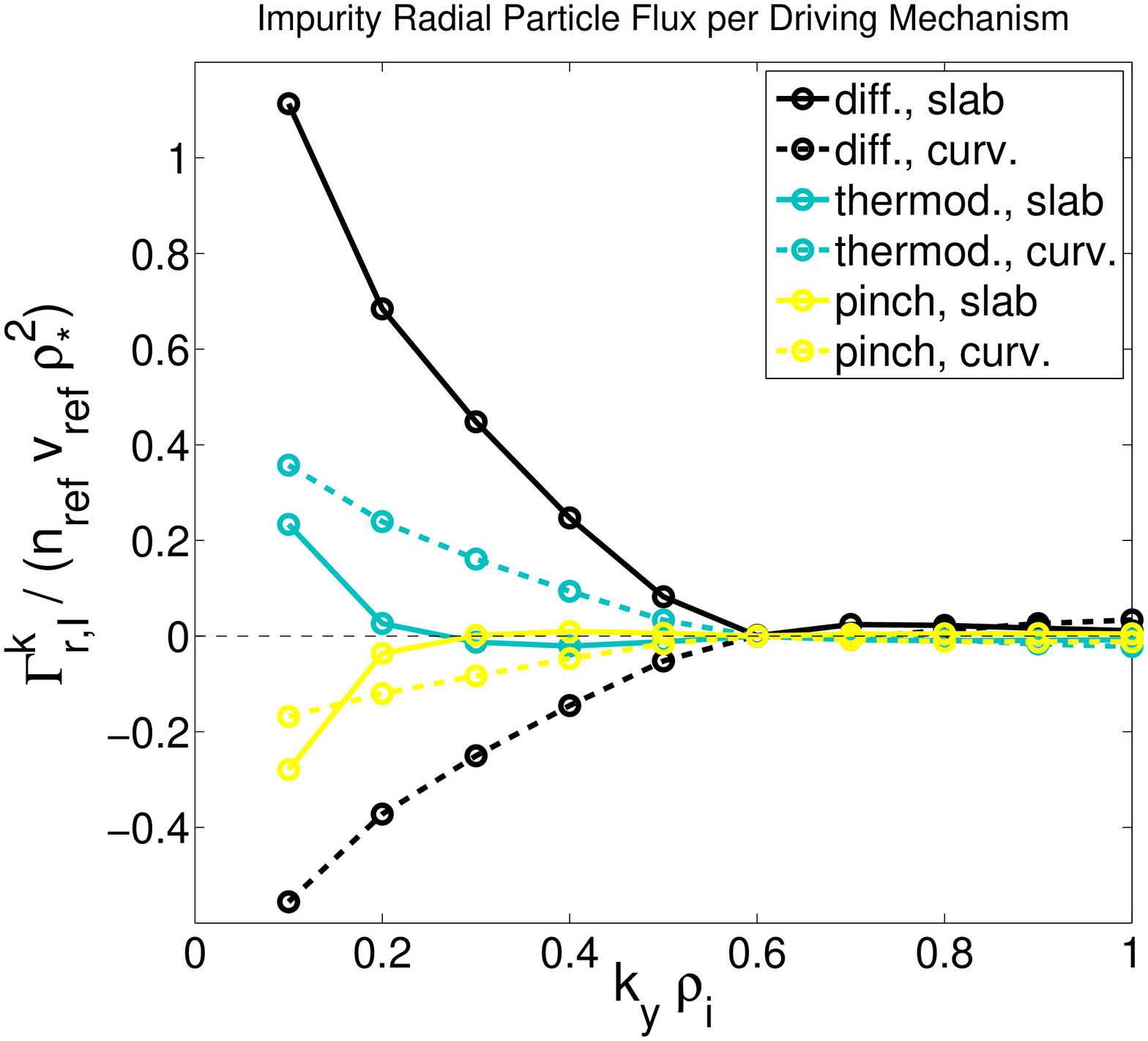}
 \end{center}
 \caption{Fluid analysis at $t=0.8 \st{s}$ with the reference $c_{\st{Li}}=0.01$ lithium concentration. Growth rates of the two most unstable modes (ITG and TEM, top left), total flux of the species driven by ITG and TE modes (top right), deuterium (bottom left) and lithium (bottom right) particle flux by slab (solid) and curvature (dashed) terms of the diffusive (black), thermo-diffusive (cyan) and pinch (yellow) contributions driven by ITG modes as a function of the bi-normal wavenumber.}
 \label{fig:fluid_qlin_t0.8}
\end{figure}

The fluid analysis of the $t=0.8 \st{s}$ case with the experimental parameters (figure \ref{fig:fluid_qlin_t0.8}) shows that ITG modes are the fastest growing eigenmodes below $k_{\st{y}} \rho_{\st{i}} \approx 0.6$. This case is characterized by strong collisionality ($\nu_{e,N}=0.46$) that stabilizes TEM-s and destabilizes drift modes rotating in the ion direction. Ion drift modes were observed also in the gyrokinetic analysis (figure \ref{fig:gk_lin_t0.8_nLi}), but, in contrast with the fluid results, they were completely stable. However, due to the inverse scaling of the flux with the bi-normal wavenumber, these modes are expected to contribute little to the overall particle transport. Electron and deuterium fluxes are similar due to the low lithium concentration, both of them are dominated by the outward slab diffusive term. 
If the lithium density is increased to $c_{\st{Li}}=15\%$ while keeping the other parameters unchanged (not shown) the ITG modes are slightly stabilized and shifted towards higher $k_{\st{y}}$ values. 
The strongest outward contribution to the radial deuterium flux is the curvature part of the diffusive term but it is compensated by thermo-diffusion, curvature pinch and an inward slab diffusion, generating a total inward deuterium transport. The structure of the lithium flux is similar to the $c_{\st{Li}}=1\%$ case with increased magnitude due to the larger impurity concentration.

\begin{figure}
 \begin{center}
  \includegraphics[scale=0.25]{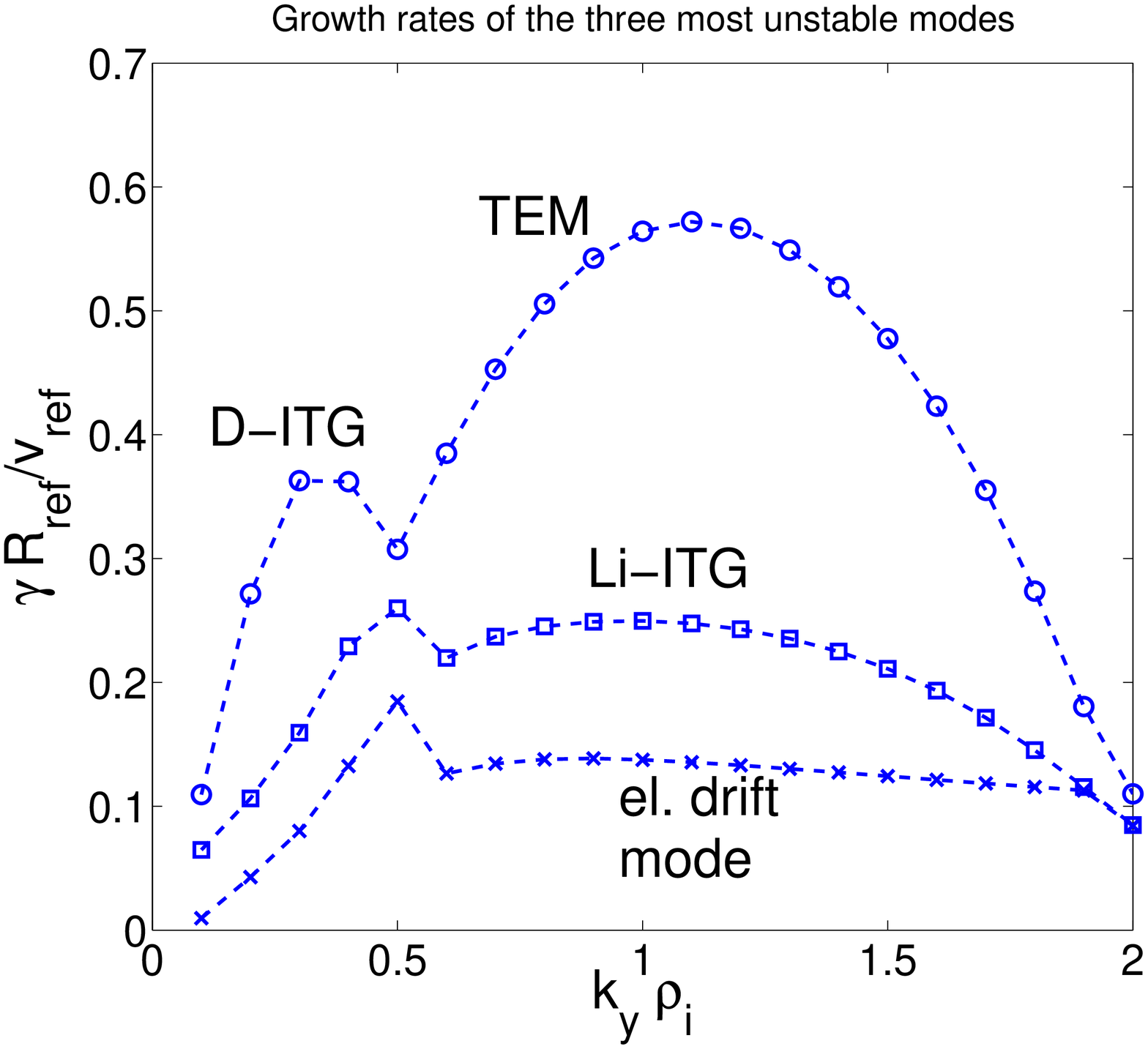}
  \includegraphics[scale=0.25]{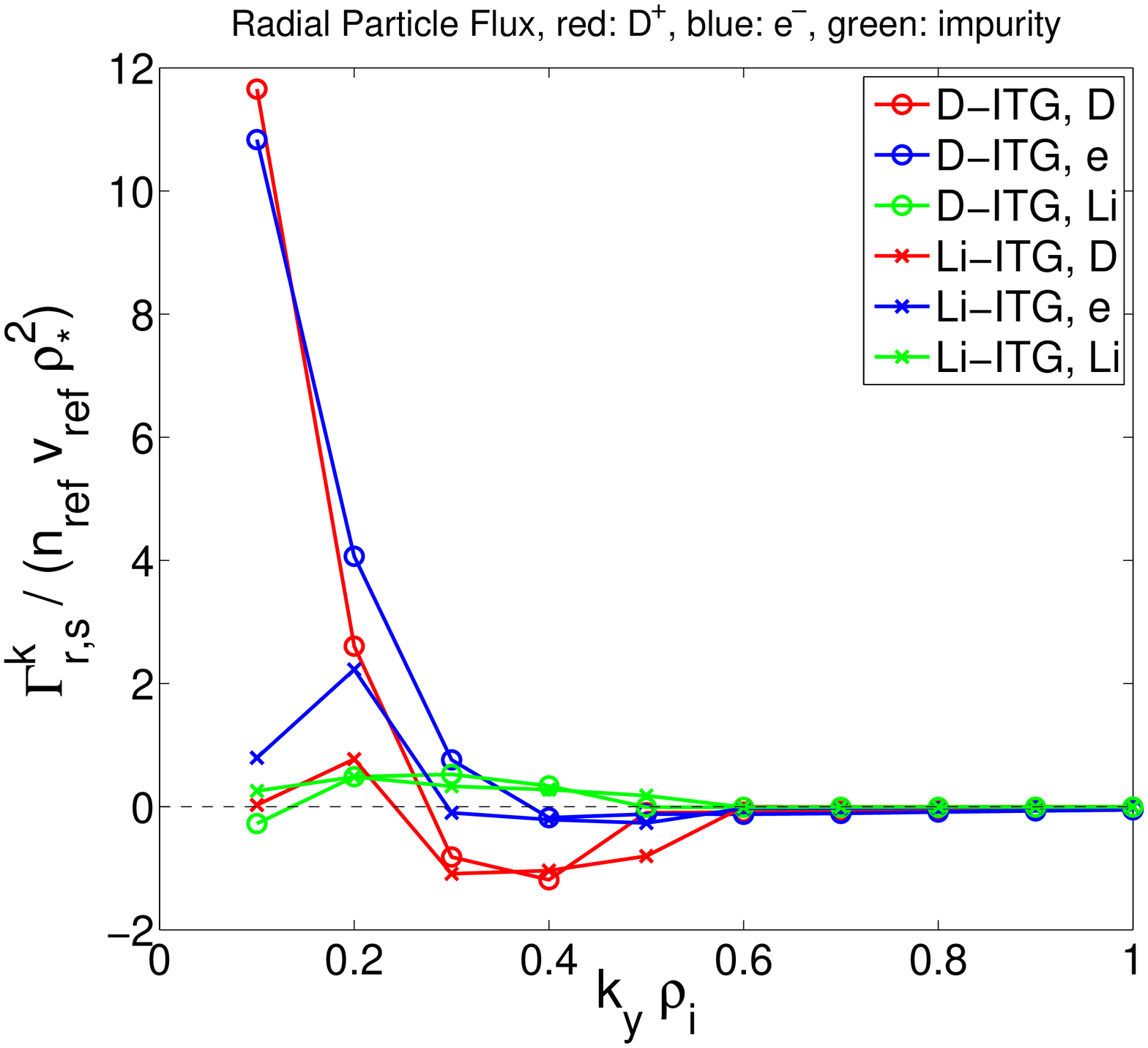} 
  \includegraphics[scale=0.25]{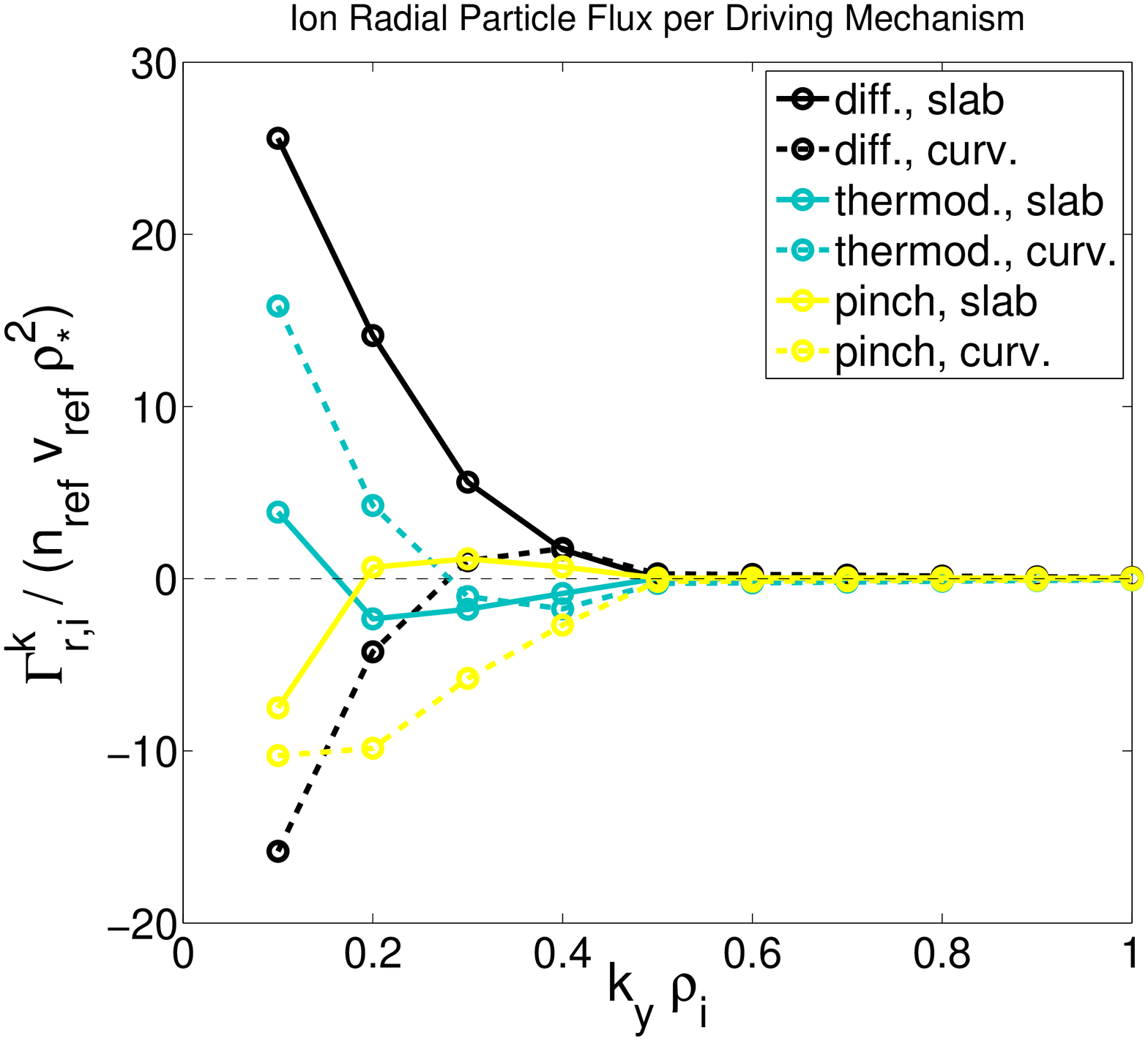}
  \includegraphics[scale=0.25]{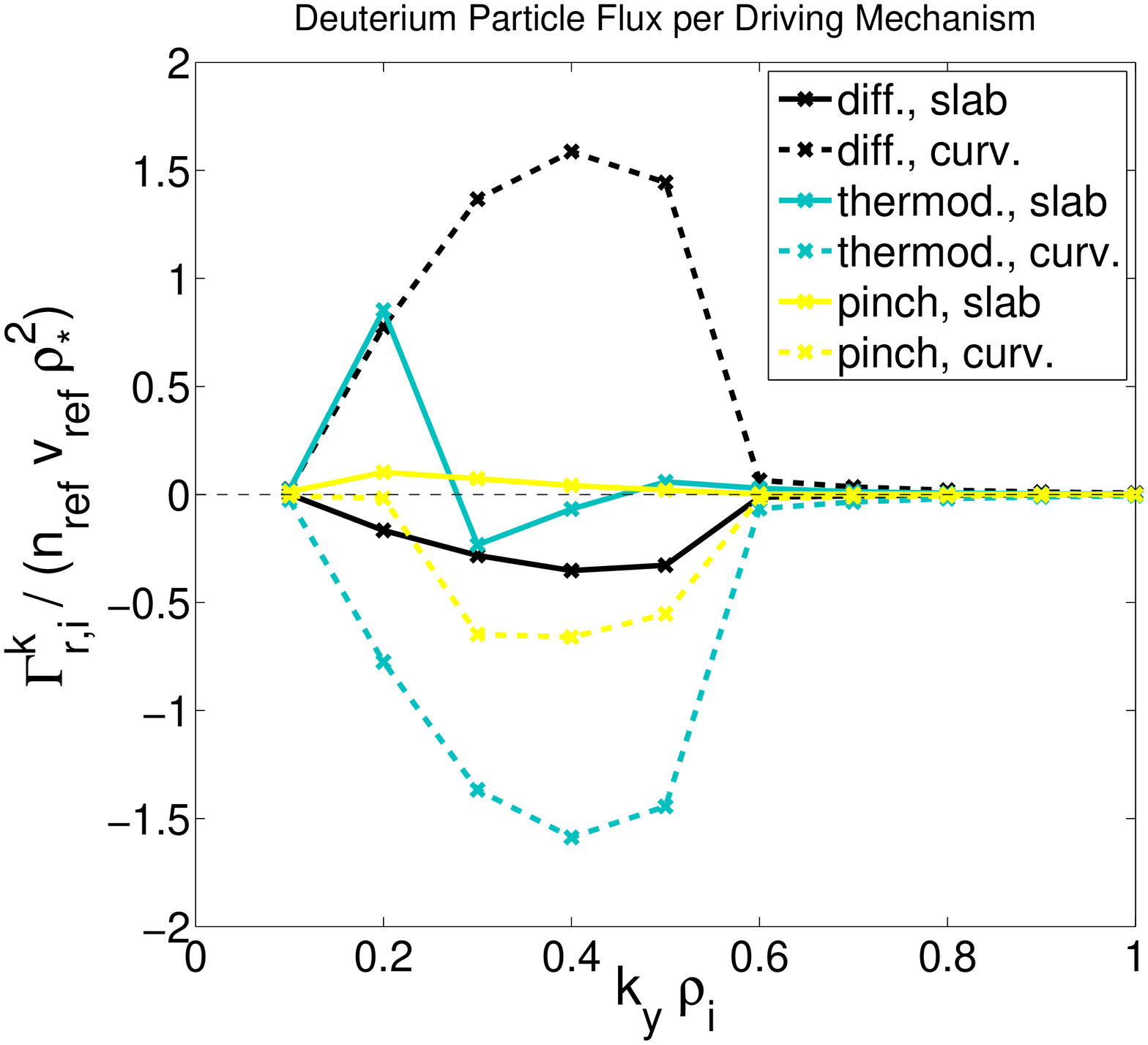}
 \end{center}
 \caption{Fluid analysis at $t=0.3 \st{s}$ with decreased $T_{\st{Li}}/T_{\st{D}}=0.5$. Growth rates of the three most unstable modes (top left), total flux of the species driven by D-ITG and Li-ITG modes (top right), deuterium particle flux by slab (solid) and curvature (dashed) terms of the diffusive (black), thermo-diffusive (cyan) and pinch (yellow) contributions driven by D-ITG (bottom left) and Li-ITG (bottom right) modes as a function of the bi-normal wavenumber.}
 \label{fig:fluid_qlin_t0.3_TLi0.5}
\end{figure}

The mixed deuterium-lithium modes in the previous cases can be split into individual eigenmodes if the Larmor-radii of the two ion species are separated. The $t=0.3 \st{s}$ case with a reduced lithium and deuterium temperature ratio $T_{\st{Li}}/T_{\st{D}}=0.5$ is shown on figure \ref{fig:fluid_qlin_t0.3_TLi0.5}. If compared to the gyrokinetic results on figure \ref{fig:gk_lin_t0.3_rhoI}, one can see that the deuterium and lithium ITG growth rates follow a similar pattern in the two models. The growth rates in the fluid calculation are slightly overestimated, probably due to the fact that only electron collisions are included, and it also predicts the presence of fast growing TE modes above $k_{\st{y}} \rho_{\st{i}}$ missing from the gyrokinetic spectrum. An important conclusion of this case is that, while we can still see the region of inward deuterium transport driven by both ITG modes between $0.2 < k_{\st{y}} \rho_{\st{i}} < 0.6$, this effect is now much less pronounced compared to the experimental case on figure \ref{fig:fluid_qlin_t0.3}. 
The bottom left and right panels show the deuterium flux distributed among the different channels, driven by the D-ITG and Li-ITG eigenmodes, respectively. 
The D-ITG modes produce a deuterium flux determined by the balance of the outward slab diffusive and inward slab thermo-diffusive and curvature pinch terms. 
The deuterium flux driven by the Li-ITG modes resemble in structure the $t=0.8 \st{s}$ case with increased lithium concentration: the slab diffusion and curvature pinch terms are directed inward while the curvature diffusion is outward. This suggests that the lithium-ITG modes, whether or not distinguished from the D-ITG as a separate eigenmode, generate a phase difference between deuterium density and potential fluctuation in a way that it drives an inward flux.

\section{Conclusions} \label{sec:conclusions}
In this work a gyrokinetic and fluid analysis of the turbulent particle transport in the FTU-LLL discharge \#30582 has been presented. Both models show that a high concentration, centrally peaked lithium density profile during the density ramp-up phase contributes to increased electron and deuterium density peaking. The presence of lithium allows the electron response to remain nearly adiabatic at the main ion scale modes, and thus to strongly reduce the electron transport. 
At the estimated lithium density gradient and concentration a dominant diffusive outward drive of the impurity flux outweighs that of the deuterium, forcing the main ions to be transported inward. 
This is achieved by the reduction of the diffusive part of the deuterium flux in presence of lithium, allowing the thermo-diffusive and pinch terms to generate a total inward flux driven by the low-k ITG modes. 

The described mechanism driving inward deuterium transport during the ramp-up phase of the discharge is not a collisional effect, it has been observed also at reduced collisionality simulations. However, the outward drive of the impurities is reinforced by strong collisionality. This indicates that light impurity seeding during the startup of a discharge might lead to improved ion density peaking also in higher temperature experiments.

The choice of lithium to achieve deuterium peaking is ideal in the sense that it is light enough to have a significant effect on the low-k ITG mode transport, and that it does not cause severe performance deterioration due to radiative losses. On the other hand, it is heavy enough that its density response sufficiently differs from that of deuterium allowing ion flow separation and leading to electron transport reduction. 

The larger gradient values, as observed following the build-up of the deuterium and electron profiles, provide stronger drive for both ITG and TE modes, and lead to a regime dominated by ion mixing modes. Increasing the gradients is therefore not sufficient to obtain the mode structure present in the density plateau phase of the discharge. The removal of the lithium impurities together with the reduction of the electron temperature are both required to obtain and ITG mode dominated turbulence in the $t=0.8 \st{s}$ case.

\ack
The authors would like to thank Clarisse Bourdelle, Sara Moradi, Geoffrey Cunningham and Peter Hill for useful discussions and ideas. This work was part-funded by RCUK Energy Programme under grant EP/I501045 and the European Communities under the contract of Association between EURATOM and CCFE. The views and opinions expressed herein do not necessarily reflect those of the European Commission. This work used resources on the HECToR supercomputer that were provided by the Engineering and Physical Sciences Research Council (EPSRC), grant number EP/H002081/1. G Szepesi's PhD is funded by EPSRC and he receives the CASE Studentship award from the Culham Centre for Fusion Energy.

\end{document}